\documentclass[prd,floatfix,onecolumn,amsmath,amssymb,floatfix]{revtex4}
\usepackage{graphicx,color,dcolumn,booktabs,bm}
\usepackage{subfigure}
\usepackage{longtable,lscape}
\usepackage{amssymb}
\usepackage{indentfirst}
\usepackage{epsfig}
\usepackage{feynmf}   
\usepackage{epstopdf}   
\usepackage{slashed}  
\usepackage{cases}
\usepackage[pdfpages]{xcolor}
\definecolor{maroon}{RGB}{139,25,150}
\usepackage{multirow}
\usepackage{float}
\usepackage{graphicx,color,dcolumn,booktabs,bm}
\usepackage[colorlinks, citecolor=blue,anchorcolor=red,menucolor=red, linkcolor=red,filecolor=red,runcolor=red,urlcolor=blue,frenchlinks=red, urlcolor=blue]{hyperref}
\usepackage{orcidlink}

\begin{document}

\preprint{}
\preprint{}
\title{\color{maroon}{Investigation of the semileptonic decays  $\Xi^{(')}_{b}\rightarrow
\Xi^{(')}_{c}{\ell}\bar\nu_{\ell}$ }}

\author{Z. Neishabouri$^{a}$\,\orcidlink{0009-0009-0892-384X}}

	\author{K.~Azizi$^{a,b}$\orcidlink{0000-0003-3741-2167}} 
	\email{kazem.azizi@ut.ac.ir} \thanks{Corresponding author}

\affiliation{
		$^{a}$Department of Physics, University of Tehran, North Karegar Avenue, Tehran 14395-547, Iran\\
		$^{b}$Department of Physics,  Dogus University, Dudullu-\"{U}mraniye, 34775 Istanbul, T\"urkiye  }
		

\date{\today}

\begin{abstract}
We study the semileptonic decays of $\Xi^{(')}_{b}\rightarrow\Xi^{(')}_{c}{\ell}\bar\nu_{\ell}$ in all  lepton channels. To do this, we first obtain the form factors defining  these decay modes within the framework of QCD sum rules. Then, using the transition form factors
 , we compute the decay widths and branching fractions for all lepton channels and compare the results of our calculations with those obtained from other theoretical methods. We  also estimate  the branching ratios and the ratio of branching fractions at different leptonic channels to provide useful information for future experiments may be planned at different Colliders.  Such comparison  will provide valuable information about the consistency/inconsistency of the SM theory predictions with experimental data in weak semileptonic single heavy baryon decays.

\end{abstract}

\maketitle
\section{Introduction}
The Standard Model (SM) is currently the most powerful framework for describing nature based on particle physics, including  three fundamental forces. However, its shortcomings in explaining phenomena such as dark matter, dark energy, gravity and unification have driven research towards exploring physics beyond the Standard Model (BSM). Following the BABAR laboratory's report of a deviation in the SM predictions regarding the Lepton Flavor Universality (LFU) in B meson decays to D mesons \cite{BaBar:2012obs}, along with results from several
other experiments on B meson decays \cite{LHCb:2014vgu,LHCb:2017vlu}, the study of these hadronic decays has gained attention as a pathway to explore BSM phenomena.  The study of weak decays of hadrons containing a bottom quark is an important tool for understanding weak transition dynamics and identifying key parameters of the Standard Model, such as elements of the Cabibbo-Kobayashi-Maskawa (CKM) matrix. It also serves to test SM predictions and search for physics beyond the Standard Model.
  In recent years, significant attention has  been given to single heavy baryons through various theoretical methods \cite{Faustov:2020thr, Ivanov:1999pz,Lyubovitskij:2003pn,Efimov:1991qj,Isgur:1991wr,Isgur:1990pm,Cho:1992cf,Le:1993gw,Farhadi:2023zeh,Kakadiya:2022zvy,Rui:2023fiz,Neishabouri:2024gbc,Azizi:2010qk,Gutsche:2018utw,Cheng:2023jpz,Majethiya:2011ry,Ghalenovi:2022adf,Hosaka:2016ypm,Duan:2024ayo} and experimental studies \cite{LHCb:2020gge,BESIII:2023ooh,LHCb:2017vhq}.
As a result, all such baryons have been observed in laboratories, with many decay channels identified  \cite{ParticleDataGroup:2024cfk}. For instance, the semileptonic decay of $\Lambda_b$ to $\Lambda_c$ has been studied using various theoretical approaches \cite{Pervin:2005ve,Faustov:2016pal,Gutsche:2014zna,Detmold:2015aaa,Gutsche:2015mxa,Miao:2022bga,Duan:2022uzm,Azizi:2018axf,Endo:2025fke}, but no deviations from the SM predictions have been observed in the LHCb experiments \cite{LHCb:2022piu}. Single heavy baryons, consisting of a heavy quark $b$ or $c$ and two light quarks, are represented in $SU(3)$ framework as $6_F\oplus\bar3_F$ states. Since baryons are fermions with half-integer total spin, they possess anti-symmetric wave functions. The color part of the wave function is a singlet state, and the spatial wave function in the ground state is symmetric. Therefore,  for a diquark with spin one, the flavor lies in the sextet $(6_F)$ representation, while for a diquark with spin zero, the flavor lies in the anti-triplet $(\bar3_F)$ representation.  
In the quark model, two baryons with the quark content $b s d (u)$ and spin-parity $J^P=\frac{1}{2}^+$ are expected, one belonging to the $6_F$ and the other to the $\bar3_F$ representation.
The first direct observation of baryon containing a heavy $b$ quark and $s$ and $d$ as light quarks, $\Xi^{-}_{b}$ occurred in $p\bar p$ collisions at $\sqrt {s} = 1.96 TeV$. This was achieved 
by reconstructing the decay $\Xi^-_b\to J/\Psi\Xi^-$ with $J/\Psi \to \mu^+\mu^-$, and $\Xi^-\to\Lambda\pi^-\to p\pi^-\pi^-$ using $D0$ detector. The signal had a significance of  $5.5\sigma$ with a measured mass of $5.774\pm0.011\pm0.015 GeV$ and spin-parity $J^P=\frac{1}{2}^+$  \cite{D0:2007gjs}. The second baryon with the same quark content, $\Xi^{'-}_b$ was observed in the LHCb collaboration and its mass difference with the $\Xi^0_b$ and $\pi^-$ has been reported as $3.653\pm0.018\pm0.006 MeV/c^2$ \cite{LHCb:2014nae}.
The primary decay channel of interest for such hadrons involving the $b$ quark decaying into a hadron with a $c$ quark,  is either semileptonically or non-leptonically.
The semileptonic decay of $\Xi_b\to\Xi_c$  has been studied using  various theoretical methods, including relativistic quark model  \cite{Faustov:2018ahb, Faustov:2020thr, Zhang:2019jax, Dutta:2018zqp, Ebert:2005ip, Ebert:2006rp, Ebert:2006hm, Ebert:2008oxa, Ivanov:1996fj}, nonralativistic quark model \cite{Albertus:2004wj,Cheng:1995fe}, light front approaches \cite{Zhao:2018zcb,Ke:2024aux,Cardarelli:1998tq,Zhao:2022vfr}, QCD sum rules (QCDSR)  \cite{Zhao:2020mod}, heavy quark effective theory (HQET)  \cite{Korner:1994nh,Boyd:1996cd}, Bethe-Salpeter approaches  \cite{Ivanov:1998ya,Rusetsky:1997id} and the spectator quark model \cite{Singleton:1990ye}.
In this work, we investigate the semileptonic decays of $\Xi^{(')}_b\to\Xi^{(')}_c$ using  QCDSR method, a powerful tool for studying weak decays in heavy baryons in non-perturbative physics. This approach relates QCD parameters like quark masses, quark and gluon condensates  (via operator product expansion and quark-hadron duality assumptions) with measurable hadronic quantities like form factors.
We calculate the form factors entering the low-energy matrix elements defining the amplitudes of the decays under study.  Our goal is to calculate decay rates and  branching ratios for all three lepton channels and compare them with predictions from other theoretical methods. Since the SM predicts identical coupling to the W and Z gauge bosons for all three lepton families, calculating branching ratios and  ratio of branching fractions  and comparing them across different leptonic channels with future experimental results provides an excellent test of SM predictions and can help identify potential deviations indicative of physics BSM like possible lepton flavor universality violation.
\\
This paper is organized within  five sections. In Sec. \ref{Sec2} the method for obtaining form factors using QCDSR is presented. Sec. \ref{Sec3}
focuses on the numerical analysis of the form factors by calculating  the working regions of the auxiliary parameters entered the calculations and finding the fit functions for all form factors in terms of $ q^2 $.  In Sec.  \ref{Sec4} the decay rates and branching ratios for all decays are computed. Section\ref{Sec5} includes concluding remarks. The Appendices provide additional details of the calculations.

\section{Method}~\label{Sec2}
The QCD sum rule method, based on the fundamental QCD Lagrangian, was proposed  by Shifman, Vanishtein and Zakharov \cite{Shifman:1978by,Shifman:1978bx}. This  approach examines hadrons in two distinct regions of the light cone:  the time-like region, in which the hadron is treated as an independent object, and the space-like region, where the dynamics of valence quarks and gluons are taken into account.
The connection between these two regions is established through dispersion integrals and the quark-hadron duality assumption, which enables hadronic parameters to be expressed in terms of QCD parameters \cite{Shifman:2001ck,Shifman:2010zzb,Gross:2022hyw,Aliev:2010uy}.
To study hadronic decays, three-point sum rules are employed to calculate form factors,
 which serve as building blocks for extracting information about decays. In this work, we focus on semileptonic decays $\Xi^{(')}_{b}\rightarrow\Xi^{(')}_{c}{\ell}\bar\nu_{\ell}$ (See Table \ref{kesi}), where the light quarks act as spectators and the $b$ quark transforms into a $c$ quark through the $W$ boson exchange via the transition current $J^{tr}=\bar c\gamma_\mu(1-\gamma_5)b$. The $W$ boson subsequently decays into leptons through a weak decay (Fig. 1).

 \begin{table}[h]
		\centering
		\begin{tabular}{|c|c|c|c|c|c|}
			\hline
			baryon\,\,\,&quark content\,\,\,&charge\,\,\,&quark model\,\,\,&spin-parity\,&mass(GeV)\\
			\hline
			$\Xi_{b}^{-}$&$(s\,d\,b)$&-1&anti-triplet&$\frac{1}{2}^{+}$&$5.797$\\
			\hline
                        $\Xi_{b}^{'-}$&$(s\,d\,b)$&-1&sextet&$\frac{1}{2}^{+}$& $5.935$\\
                        \hline
                        $\Xi_{b}^{0}$&$(s\,u\,b)$&0&anti-triplet&$\frac{1}{2}^{+}$& $5.791$\\
			\hline
                        $\Xi_{b}^{'0}$&$(s\,u\,b)$&0&sextet&$\frac{1}{2}^{+}$& -\\
			\hline
		\end{tabular}
		\caption{ 
			Quantum numbers and quark content of\,\,$ \Xi_{b}$}\label{kesi}
	\end{table}

\begin{figure}[h!] 
\includegraphics[totalheight=5.8cm,width=5.9cm]{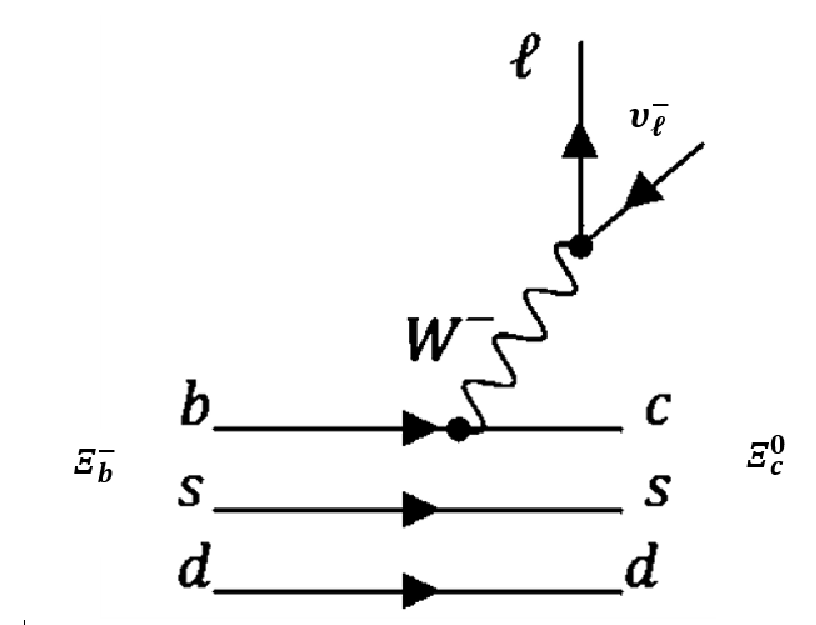}
\caption{Schematic of  $\Xi^{-}_{b}\rightarrow\Xi^{0}_{c}{\ell}\bar\nu_{\ell}$ decays }\label{kesi}
\end{figure}
The three-point correlation function in QCDSR is expressed as:
\begin{eqnarray}\label{CorFunc}
\Pi_{\mu}(p,p^{\prime},q)&=&i^2\int d^{4}x e^{-ip\cdot x}\int d^{4}y e^{ip'\cdot y}  \langle 0|{\cal T}\{{\cal J}^{\Xi^{(')}_c}(y){\cal
J}_{\mu}^{tr,V(A)}(0) \bar{\cal J}^{\Xi^{(')}_b}(x)\}|0\rangle,
\end{eqnarray}
where  $ {\cal J}^{\Xi_{b}}(x)$ and ${\cal J}^{\Xi_{c}}(y) $ are initial and final hadron currents, respectively,  ${\cal J}^{tr}$ is the transition current for weak decays and $\cal T$ is the time ordering operator.
	
\subsection{Phenomenological side}
The effective Hamiltonian corresponding to the decay is:
\begin{eqnarray}\label{Heff}
{\cal H}_{eff} =
\frac{G_F}{\sqrt2} V_{cb} ~\bar c \gamma_\mu(1-\gamma_5) b ~\bar{\ell}\gamma^\mu(1-\gamma_5) \nu_{\ell},
\end{eqnarray}
where $G_F$ is the Fermi coupling constant and $V_{cb}$ is the CKM matrix element. The decay amplitude is obtained by sandwiching ${\cal H}_{eff}$ between the initial and final baryon states:
\begin{eqnarray}\label{amp}
&&M=\frac{G_F}{\sqrt2} V_{cb}\bar{\ell}~\gamma^\mu(1-\gamma_5) \nu_{\ell} \langle \Xi_{c}\vert \bar c \gamma_\mu(1-\gamma_5)b\vert \Xi_{b}\rangle.
\end{eqnarray}
The weak decay involves vector ($V^\mu$) and the axial vector ($A^\mu$) transitions.  The matrix elements are parametrized using Lorentz invariance and parity:
\begin{eqnarray}\label{Cur.with FormFac.}
\mbox{Vector part:}
&&\langle \Xi_c(p',s')|V^{\mu}|\Xi_b (p,s)\rangle = \bar
u_{\Xi_c}(p',s') \Big[F_1(q^2)\gamma^{\mu}+F_2(q^2)\frac{p^{\mu}}{m_{\Xi_b}}
+F_3(q^2)\frac{p'^{\mu}}{m_{\Xi_c}}\Big] u_{\Xi_b}(p,s), \notag \\
\mbox{Axial-vector part:}
&&\langle \Xi_c(p',s')|A^{\mu}|\Xi_b (p,s)\rangle = \bar u_{\Xi_c}(p',s') \Big[G_1(q^2)\gamma^{\mu}+G_2(q^2)\frac{p^{\mu}}{m_{\Xi_b}}+G_3(q^2)\frac{p'^{\mu}}{m_{\Xi_c}}\Big]
\gamma_5 u_{\Xi_b}(p,s). \notag \\
\end{eqnarray}
Here,  $ F_i(q^2)$ and $ G_i(q^2) ~(i=1,2,3) $ are form factors that encode non-peturbative QCD effects. The momentum transfer to leptons is $q=p-p'$, and the Dirac spinors are denoted by $u_{\Xi_{b}}(p,s)$ for initial baryon and $u_{\Xi_{c}}(p',s')$ for final baryonic state. 
\\
Now, to obtain the phenomenological side (hadronic side) of the correlation function,  we insert two complete sets corresponding to the quantum numbers of the initial and final baryons into Eq.  \ref{CorFunc} \cite{Colangelo:2000dp}, 
\begin{eqnarray} \label{compelet set}
1=\vert 0\rangle\langle0\vert +\sum_{h}\int\frac{d^4p}{(2\pi)^4}(2\pi) \delta(p^2-m^2_h)|h(p)\rangle  \langle h(p)|+\mbox{higher Fock states},
\end{eqnarray}
where $h(p)$ is the possible hadronic state with momentum $p$ and mass $m_h$.
\\
After Fourier transformation and integration over space-time coordinates ($x,y$) the hadronic representation of the correlation function becomes:

\begin{eqnarray} \label{PhysSide}
\Pi_{\mu}^{Phys.}(p,p',q)=\frac{\langle 0 \mid {\cal J}^{\Xi_c} (0)\mid \Xi_c(p') \rangle \langle \Xi_{c} (p')\mid
{\cal J}_{\mu}^{tr,V(A)}(0)\mid \Xi_b(p) \rangle \langle \Xi_{b}(p)
\mid \bar {\cal J}^{\Xi_b}(0)\mid
0\rangle}{(p'^2-m_{\Xi_c}^2)(p^2-m_{\Xi_b}^2)}+\cdots~,
\end{eqnarray}
where higher states contribute to additional terms. $\lambda_{\Xi_b}$ and $\lambda_{\Xi_c}$ are defined as the residue of the initial and final states:
\begin{eqnarray}\label{MatrixElements}
&&\langle 0|{\cal J}^{\Xi_c}(0)|\Xi_c(p')\rangle =
\lambda_{\Xi_c} u_{\Xi_c}(p',s'), \notag \\
&&\langle\Xi_b(p)|\bar {\cal J}^{\Xi_b}(0)| 0 \rangle =
\lambda^{+}_{\Xi_b}\bar u_{\Xi_b}(p,s).
\end{eqnarray}
The Dirac spinors satisfy the following summation relations:
\begin{eqnarray}\label{Spinors}
\sum_{s'} u_{\Xi_c} (p',s')~\bar{u}_{\Xi_c}
(p',s')&=&\slashed{p}~'+m_{\Xi_c},\notag \\
\sum_{s} u_{\Xi_b}(p,s)~\bar{u}_{\Xi_b}(p,s)&=&\slashed
p+m_{\Xi_b}.
\end{eqnarray}
Now, we insert all the matrix elements, defined above into Eq.  (\ref{PhysSide}), and perform the double Borel transformation to suppress the contributions from the higher states and continuum. A double Borel transformation is applied to the correlation function with introducing auxiliary parameters $ M^2 $ and $M'^{2}$, which are chosen in numerical analysis to ensure stability. The transformation is expressed as \cite{Aliev:2006gk}:
\begin{eqnarray}\label{BorelQCD2}
\mathbf{\widehat{B}}\frac{1}{(p^{2}-s)^m} \frac{1}{(p'^{2}-s')^n}\longrightarrow (-1)^{m+n}\frac{1}{\Gamma[m]\Gamma[n]} \frac{1}{(M^2)^{m-1}}\frac{1}{(M'^2)^{n-1}}e^{-s/M^2} e^{-s'/M'^2}.
\end{eqnarray}

After applying the double Borel transformation, the phenomenological side of the correlation function becomes:
\begin{eqnarray}\label{Physical Side structures}
&&\mathbf{\widehat{B}}~\Pi_{\mu}^{\mathrm{Phys.}}(p,p',q)=\lambda_{\Xi_b}\lambda_{\Xi_c}~e^{-\frac{m_{\Xi_b}^2}{M^2}}
~e^{-\frac{m_{\Xi_c}^2}{M'^{2}}}\Bigg[F_{1}\bigg(m_{\Xi_b} m_{\Xi_c} \gamma_{\mu}+m_{\Xi_b} \slashed{p}' \gamma_{\mu}+m_{\Xi_c}\gamma_{\mu}\slashed {p}+\slashed {p}'\gamma_\mu\slashed {p}\bigg)+\notag\\
&&F_2\bigg(\frac{m_{\Xi_c}}{m_{\Xi_b}}p_\mu\slashed {p}+\frac{1}{m_{\Xi_b}}p_{\mu}\slashed {p}' \slashed {p}+m_{\Xi_c}p_\mu +p_\mu\slashed {p}'\bigg)+ F_3\bigg(\frac{1}{m_{\Xi_c}} p'_{\mu} \slashed {p}' \slashed{p}+p'_\mu\slashed {p}'+p'_\mu\slashed {p}+m_{\Xi_b}p'_\mu\bigg)-\notag\\
&& G_1\bigg(m_{\Xi_b} m_{\Xi_c} \gamma_{\mu}\gamma_{5}+m_{\Xi_b}\slashed {p}'\gamma_\mu\gamma_5-m_{\Xi_c}\gamma_\mu\slashed {p}\gamma_5-\slashed {p}'\gamma_\mu\slashed {p}\gamma_5\bigg)- G_2\bigg(p_\mu\slashed {p}'\gamma_5+m_{\Xi_c}p_\mu\gamma_5-\frac{m_{\Xi_c}}{m_{\Xi_b}}p_\mu\slashed {p}\gamma_5-\frac{1}{m_{\Xi_b}} p_{\mu} \slashed {p}' \slashed
{p}\gamma_{5}\bigg)-\notag\\
&&G_3\bigg(\frac{m_{\Xi_b}}{m_{\Xi_c}}p'_\mu\slashed {p}'\gamma_5+m_{\Xi_b}p'_\mu\gamma_5-\frac{1}{m_{\Xi_c}} p'_{\mu}
\slashed {p}'\slashed{p}\gamma_{5}-p'_\mu\slashed {p}\gamma_5\bigg)\Bigg]+..., \notag \\
\end{eqnarray}
\\
\subsection{QCD side}
To obtain the QCD side of the correlation function in the deep Euclidean region, an operator product expansion (OPE) is performed.
The time-ordered product of two currents at different points can be expanded in terms of local operators.
\begin{eqnarray}\label{OPE}
{\cal T}{\{j(x)\bar j(0)\}}=\sum_d C_d(x^2) O_d,
\end{eqnarray}
where $C_d$ are Wilson coefficients and $O_d$ denote a set of local operators ordered {\color{red}by} their dimensions. In the present study, this formalism {\color{red}is applied}to the three-current correlation function.
\\
The first step to drive the correlation function in an OPE form is inserting the currents of the initial and final baryons in terms of their quark content in Eq.\ (\ref{CorFunc}).
The interpolating current of single heavy $\Xi^{(')}_Q$ baryon with spin-parity $J^P=(\frac{1}{2})^+$ is given by \cite{Azizi:2011mw}:
%
\begin{eqnarray}\label{current}
&&J^{\Xi_Q} =\frac {1}{ \sqrt{6}} \epsilon^{abc} \Bigg\{2\Big( q_1^{aT} C q_2^b \Big) \gamma_5 Q^c + 2 \beta \Big( q_1^{aT}
C \gamma_5 q_2^b \Big) Q^c + \Big( q_1^{aT} C Q^b \Big) \gamma_5 q_2^c +\beta \Big(q_1^{aT} C\gamma_5 Q^b \Big) q_2^c +\notag\\
 &&
\Big(Q^{aT} C q_2^b \Big) \gamma_5 q_1^c + \beta \Big(Q^{aT} C
\gamma_5 q_2^b \Big) q_1^c \Bigg\}~,\notag\\
&& J^{\Xi'_Q}= -\frac {1}{ \sqrt{2}} \epsilon^{abc} \Bigg\{\Big(q_1^{aT} C Q^b \Big) \gamma_5 q_2^c + \beta \Big( q_1^{aT} C
\gamma_5 Q^b \Big) q_2^c - \Big[\Big( Q^{aT} C q_2^b \Big) \gamma_5 q_1^c + \beta \Big( Q^{aT} C \gamma_5 q_2^b \Big) q_1^c \Big] \Bigg\},
\end{eqnarray}
where $a,~b$ and $c$ are color indices, $C$ is the charge
conjugation operator, $q_1$ and $q_2$ are the light quarks and $Q$ is bottom or charm 
quark field. The $\beta$ is an auxiliary parameter that allows to consider all possible arrangements of quarks with $\beta=-1$
being corresponding to  Ioffe current. Its region should be fixed in the  numerical analysis section.
After inserting the interpolating currents of the initial and final baryons $({\cal J}^{\Xi_{b}}, {\cal J}^{\Xi_{c}})$ and the transition current $(J^{tr})$ into the correlation function Eq.  (\ref{CorFunc}), and utilizing the Wick theorem to account for all possible quark contractions, the QCD side of correlation function for the semileptonic decay $\Xi^{-}_{b}\rightarrow\Xi^{0}_{c}{\ell}\bar\nu_{\ell}$ takes the following form [correlation functions of other decays are given in the Appendix A]:
\begin{eqnarray} \label{36 term}
&&\Pi^{QCD}_{\mu}(p,p',q)=i^2 \int d^4x e^{-ipx}\int d^4y e^{ip'y} \frac{1}{6} \epsilon_{a'b'c'} \epsilon_{abc}\Bigg\{4Tr [S'^{aa'}_d(y-x) S^{bb'}_s(y-x)]\gamma_5 S^{ci}_c(y) \gamma_\mu (1-\gamma_5) S^{ic'}(-x)_b\gamma_5\notag\\
&&-4\beta Tr [\gamma_5 S'^{aa'}_d(y-x) S^{bb'}_s(y-x)] \gamma_5 S^{ci}_c(y) \gamma_\mu (1-\gamma_5) S^{ic'}(-x)_b-2\gamma_5 S^{ci}_c(y)\gamma_\mu(1-\gamma_5) S^{ib'}_b(-x) S'^{aa'}_d(y-x) S^{bc'}_s(y-x) \gamma_5\notag\\
&&+2\beta \gamma_5 S^{ci}_c(y)\gamma_\mu(1-\gamma_5) S^{ib'}_b(-x)\gamma_5 S'^{aa'}_d(y-x) S^{bc'}_s(y-x) -2\gamma_5 S^{ci}_c(y)\gamma_\mu(1-\gamma_5) S^{ia'}_b(-x) S'^{bb'}_s(y-x)S^{ac'}_d(y-x)\gamma_5\notag\\
&&+2\beta \gamma_5 S^{ci}_c(y)\gamma_\mu(1-\gamma_5) S^{ia'}_b(-x)\gamma_5 S'^{bb'}_s(y-x) S^{ac'}_d(y-x)+4\beta Tr [S'^{aa'}_d(y-x)\gamma_5 S^{bb'}_s(y-x)] S^{ci}_c(y) \gamma_\mu(1-\gamma_5) S^{icc}_b(-x) \gamma_5\notag\\
&&-4\beta ^2 Tr [\gamma_5 S'^{aa'}_d(y-x)\gamma_5 S^{bb'}_s(y-x)] S^{ci}_c(y) \gamma_\mu(1-\gamma_5) S^{icc}_b(-x)-2\beta S^{ci}_c(y) \gamma_\mu(1-\gamma_5) S^{ib'}_b(-x) S'^{aa'}_d(y-x) \gamma_5 S^{bc'}_s(y-x)\gamma_5\notag\\
&&+2\beta^2 S^{ci}_c(y) \gamma_\mu(1-\gamma_5) S^{ib'}_b(-x) \gamma_5 S'^{aa'}_d(y-x) \gamma_5 S^{bc'}_s(y-x)-2\beta S^{ci}_c(y) \gamma_\mu(1-\gamma_5) S^{ia'}_b(-x) S'^{bb'}_s(y-x) \gamma_5 S^{ac'}_d(y-x)\gamma_5\notag\\
&&+2\beta^2 S^{ci}_c(y) \gamma_\mu(1-\gamma_5) S^{ia'}_b(-x)\gamma_5 S'^{bb'}_s(y-x) \gamma_5 S^{ac'}_d(y-x)-2\gamma_5 S^{cb'}_s(y-x) S'^{aa'}_d(y-x) S^{bi}_c(y)\gamma_\mu(1-\gamma_5) S^{ic'}_b(-x)\gamma_5\notag\\
&&+2\beta\gamma_5 S^{cb'}_s(y-x)\gamma_5 S'^{aa'}_d(y-x) S^{bi}_c(y)\gamma_\mu(1-\gamma_5) S^{ic'}_b(-x)+Tr[S'^{aa'}_d(y-x) S^{bi}_c(y)\gamma_\mu(1-\gamma_5)S^{ib'}_b(-x)] \gamma_5 S^{cc'}_s(y-x)\gamma_5\notag\\
&&-\beta Tr[\gamma_5 S'^{aa'}_d(y-x) S^{bi}_c(y)\gamma_\mu(1-\gamma_5)S^{ib'}_b(-x)] \gamma_5 S^{cc'}_s(y-x)-\gamma_5 S^{cb'}_s(y-x) S'^{ia'}_b(-x)(1-\gamma_5)\gamma_\mu S'^{bi}_c(y) S^{ac'}_d(y-x)\gamma_5\notag\\
&&+\gamma_5 S^{cb'}_s(y-x) \gamma_5 S'^{ia'}_b(-x)(1-\gamma_5)\gamma_\mu S'^{bi}_c(y) S^{ac'}_d(y-x)-2\beta S^{cb'}_s(y-x) S'^{aa'}_d(y-x)\gamma_5 S^{bi}_c(y) \gamma_\mu(1-\gamma_5) S^{ic'}_b(-x)\gamma_5\notag\\
&&+2\beta^2 S^{cb'}_s(y-x)\gamma_5 S'^{aa'}_d(y-x)\gamma_5 S^{bi}_c(y) \gamma_\mu(1-\gamma_5) S^{ic'}_b(-x)+\beta Tr[S'^{aa'}_d(y-x)\gamma_5 S^{bi}_c(y)\gamma_\mu(1-\gamma_5) S^{ib'}_b(-x)] S^{cc'}_s(y-x)\gamma_5\notag\\
&&-\beta^2 Tr[\gamma_5 S'^{aa'}_d(y-x)\gamma_5 S^{bi}_c(y)\gamma_\mu(1-\gamma_5) S^{ib'}_b(-x)] S^{cc'}_s(y-x)+\beta S^{cb'}_s(y-x) S'^{ia'}_b(-x) (1-\gamma_5)\gamma_\mu S'^{bi}_c(y)\gamma_5 S^{ac'}_d(y-x)\gamma_5\notag\\
&&-\beta^2 S^{cb'}_s(y-x) \gamma_5 S'^{ia'}_b(-x) (1-\gamma_5)\gamma_\mu S'^{bi}_c(y)\gamma_5 S^{ac'}_d(y-x)-2\gamma_5 S^{ca'}_d(y-x) S'^{bb'}_s(y-x) S^{ai}_c(y)\gamma_\mu(1-\gamma_5) S^{ic'}_b(-x)\gamma_5\notag\\
&&+2\beta \gamma_5 S^{ca'}_d(y-x)\gamma_5 S'^{bb'}_s(y-x) S^{ai}_c(y)\gamma_\mu(1-\gamma_5) S^{ic'}_b(-x)-\gamma_5 S^{ca'}_d(y-x) S'^{ib'}_b(-x) (1-\gamma_5)\gamma_\mu S'^{ai}_c(y) S^{bc'}_s(y-x)\gamma_5\notag\\
&&+\beta \gamma_5 S^{ca'}_d(y-x) \gamma_5 S'^{ib'}_b(-x) (1-\gamma_5)\gamma_\mu S'^{ai}_c(y) S^{bc'}_s(y-x)+Tr[S'^{ia'}_b(-x)(1-\gamma_5)\gamma_\mu S'^{ai}_c(y) S^{bb'}_s(y-x)]\gamma_5 S^{cc'}_d(y-x)\gamma_5\notag\\
&&-\beta Tr[\gamma_5S'^{ia'}_b(-x)(1-\gamma_5)\gamma_\mu S'^{ai}_c(y) S^{bb'}_s(y-x)]\gamma_5 S^{cc'}_d(y-x)-2\beta S^{ca'}_d(y-x) S'^{bb'}_s(y-x)\gamma_5 S^{ai}_c(y) \gamma_\mu(1-\gamma_5) S^{ic'}_b(-x)\gamma_5\notag\\
&&+2\beta^2  S^{ca'}_d(y-x) \gamma_5 S'^{bb'}_s(y-x)\gamma_5 S^{ai}_c(y) \gamma_\mu(1-\gamma_5) S^{ic'}_b(-x)-\beta S^{ca'}_d(y-x) S'^{ib'}_b(-x) (1-\gamma_5)\gamma_\mu S'^{ai}_c(y)\gamma_5 S^{bc'}_s(y-x)\gamma_5\notag\\
&&+\beta^2 S^{ca'}_d(y-x) \gamma_5 S'^{ib'}_b(-x) (1-\gamma_5)\gamma_\mu S'^{ai}_c(y)\gamma_5 S^{bc'}_s(y-x)+\beta Tr[S^{ai}_c(y)\gamma_\mu(1-\gamma_5) S^{ia'}_b(-x) S'^{bb'}_s(y-x)\gamma_ 5] S^{cc'}_d(y-x)\gamma_5\notag\\
&&-\beta^2 Tr[S^{ai}_c(y)\gamma_\mu(1-\gamma_5) S^{ia'}_b(-x)\gamma_5 S'^{bb'}_s(y-x)\gamma_5] S^{cc'}_d(y-x)
\Bigg\},
\end{eqnarray}
where $S_q$ are the light quark propagators,  $S_Q$ are the heavy propagators and $S'_q=C S^T C$.
The light quark propagator is given by \cite{Agaev:2020zad}:
\begin{eqnarray}\label{LightProp}
&&S_{q}^{ab}(x)=i\delta _{ab}\frac{\slashed x}{2\pi ^{2}x^{4}}-\delta _{ab}%
\frac{m_{q}}{4\pi ^{2}x^{2}}-\delta _{ab}\frac{\langle\overline{q}q\rangle}{12} +i\delta _{ab}\frac{\slashed xm_{q}\langle \overline{q}q\rangle }{48}%
-\delta _{ab}\frac{x^{2}}{192}\langle \overline{q}g_{}\sigma
Gq\rangle+\notag\\
&&i\delta _{ab}\frac{x^{2}\slashed xm_{q}}{1152}\langle \overline{q}g_{}\sigma Gq\rangle-i\frac{g_{}G_{ab}^{\alpha \beta }}{32\pi ^{2}x^{2}}\left[ \slashed x{\sigma _{\alpha \beta }+\sigma _{\alpha \beta }}\slashed x\right]-i\delta _{ab}\frac{x^{2}\slashed xg_{}^{2}\langle
\overline{q}q\rangle ^{2}}{7776} -\delta _{ab}\frac{x^{4}\langle \overline{q}q\rangle \langle
g_{}^{2}G^{2}\rangle }{27648}+\ldots,
\end{eqnarray}
and the heavy quark propagator is represented as  \cite{Agaev:2020zad}:
\begin{eqnarray}\label{HeavyProp}
&&S_{Q}^{ab}(x)=i\int \frac{d^{4}k}{(2\pi )^{4}}e^{-ikx}\Bigg
\{\frac{\delta_{ab}\left( {\slashed k}+m_{Q}\right) }{k^{2}-m_{Q}^{2}}-\frac{g_{}G_{ab}^{\mu \nu}}{4}\frac{\sigma _{\mu\nu }\left( {%
\slashed k}+m_{Q}\right) +\left( {\slashed k}+m_{Q}\right) \sigma
_{\mu\nu}}{(k^{2}-m_{Q}^{2})^{2}} +\frac{g_{}^{2}G^{2}}{12}\delta _{ab}m_{Q}\frac{k^{2}+m_{Q}{\slashed k}}{%
(k^{2}-m_{Q}^{2})^{4}}\notag\\
&&+\ldots\Bigg \},
\end{eqnarray}

where 
\begin{eqnarray}\label{GluonField}
&&G_{ab}^{\mu \nu }=G_{A}^{\mu\nu
}t_{ab}^{A},\,\,~~G^{2}=G_{A}^{\mu\nu} G_{\mu \nu }^{A},~~ t^{A}=\lambda ^{A}/2,
\end{eqnarray}
$G_{\mu\nu}$ is the gluon field strength tensor, $\lambda ^{A}$ are  the Gell-Mann matrices (where $A=1,\,2\,\ldots 8$); $\mu$ and $\nu$ are Lorentz indices. Each term in quark propagator is an operator with a specific mass dimension. The OPE consists of both perturbative contributions (e.g., the bare loop, d=0) and non-perturbative contributions such as: $d=3, \langle \overline{q}q\rangle$, $d=4, \langle G^2\rangle$, $d=5, \langle \overline{q}g_{}\sigma Gq\rangle$, $d=6, \langle\overline{q}q\rangle^2$ and higher-order corrections form. By substituting the quark propagators into the correlation function on the QCD side,  Eq.\ref{36 term},  both the perturbative and non-perturbative contributions up to dimension six are obtained. Through mathematical calculations involving Fourier integrals, four-momentum integrals, various identities, the Dirac delta function and Feynman parametrization,  the QCD side correlation function is derived. This function includes twenty-four Lorentz structures multiplied by invariant functions. To illustrate,  the details of the calculations are presented in the Appendix B.

The QCD side of the correlation function is expressed as:
\begin{eqnarray}\label{Structures}
\Pi_{\mu}^{\mathrm{QCD}}(p,p',q)=\Sigma_i\Pi^{\mathrm{QCD}}_i (p^2,p'^2,q^2) S_i,
\end{eqnarray}
where $S_i$ represent the 24 Lorentz structures:\\ $p_\mu, p'_\mu, \gamma_\mu, \gamma_\mu\gamma_5, p_\mu\gamma_5, p'_\mu\gamma_5, p_\mu \slashed {p}, p'_\mu\slashed {p},p_\mu\slashed {p}',p'_\mu\slashed {p}', p'_\mu\slashed {p}\gamma_5, p_\mu\slashed {p}\gamma_5, \gamma_\mu\slashed {p},\slashed {p}'\gamma_\mu, \gamma_\mu\slashed {p}\gamma_5,
\slashed {p}'\gamma_\mu\slashed {p}, p_\mu\slashed {p}'\slashed {p}, p'_\mu\slashed {p}'\slashed {p}, p'_\mu\slashed {p}'\gamma_5, p_\mu\slashed {p}'\gamma_5,\\
 p'_\mu\slashed {p}'\slashed {p}\gamma_5, p_\mu\slashed{p}'\slashed{p}\gamma_5, \slashed {p}'\gamma_\mu\gamma_5$ and $\slashed {p}'\gamma_\mu\slashed {p}\gamma_5$.
The invariant functions $\Pi^{\mathrm{QCD}}_i(p^{2},p'^{2},q^{2})$ ($i$ related to different structures) are defined via double dispersion integrals:
\begin{eqnarray}\label{PiQCD}
\Pi^{\mathrm{QCD}}_i(p^{2},p'^{2},q^{2})&=&\int_{s_{min}}^{\infty}ds
\int_{s'_{min}}^{\infty}ds'~\frac{\rho
^{\mathrm{QCD}}_i(s,s',q^{2})}{(s-p^{2})(s'-p'^{2})} \notag\\
&+&\Gamma_i(p^2,p'^2,q^2),
\end{eqnarray}
where, $s_{min}=(m_d+m_s+m_b)^{2}$ and $s'_{min}=(m_d+m_s+m_c)^{2}$. On the other hand, using the identity leads to the presence of negative logarithmic functions (Eq. \ref{gamma}) which can be defined by imaginary parts;  $\rho_i^{\mathrm{QCD}}(s,s',q^{2})=\frac{1}{\pi}Im\Pi^{QCD}_i(p^2,p'^2,q^2)$ where  $\rho_i^{\mathrm{QCD}}(s,s',q^{2})$ are called spectral densities. In addition, in the fifth and sixth dimensions of the mass, there are terms that lack the imaginary parts and are calculated directly that are displayed as $\Gamma_i(p^2,p'^2,q^2)$  in Eq.  (\ref{PiQCD}). 
Therefore, the spectral densities have two contributions, perturbative part ($\rho_i ^{Pert.}(s,s',q^{2})$) and non-perturbative parts for the quark condensates ($n=3$) and the gluon condensates  ($n= 4$). After implementing quark-hadron duality and converting the upper limits of the integrals to $s_0$ and $s'_0$, which are continuum thresholds at the initial and final states, respectively and performing the Borel transformation to subtract the contributions of the higher resonances and continuum we get the invariant functions of the QCD side:

\begin{eqnarray}\label{qcd part2}
&&\Pi^{\mathrm{QCD}}_i (M^2,M'^2,s_0,s'_0,q^2)=\int _{s_{min}}^{s_0} ds\int _{s'_{min}}^{s'_0}ds' e^{-s/M^2} e^{-s'/M'^2} \Big[\rho_i ^{Pert.}(s,s',q^{2})+\sum_{n=3}^{4}\rho_{i}^{n}(s,s',q^{2})\Big]+\notag\\
&&\mathbf{\widehat{B}}\Big[\Gamma_i(p^{2},p'^{2},q^{2})\Big].\nonumber\\
\end{eqnarray}
The components of  $\rho_{i}(s,s^{\prime},q^2)$ and $\Gamma_i(p^{2},p'^{2},q^{2})$ are given, as an example for the structure $p_{\mu}\slashed {p'}\slashed {p}$,  in Appendix C.
 
Finally, by matching the QCD side correlation function  Eq.  \ref{Structures} with its physical counterpart  Eq.  \ref{Physical Side structures} and equating corresponding Lorentz structures on both sides yield the form factors in terms of  QCD degrees of freedom and other related parameters such as auxiliary ones entered the calculations.  This process provides a robust framework for extracting non-perturbative QCD information and calculating decay observables like branching ratios.
 In the next section,  the form factors are numerically analyzed by determining   the range of the auxiliary parameters.

\section{Numerical Analysis}\label{Sec3}
To calculate the decay rates of  $\Xi^{(')}_{b}\rightarrow\Xi^{(')}_{c}{\ell}\bar\nu_{\ell}$,  obtaining the $ q^2 $-dependency of  form factors is a crucial step. Since the sum rules for these form factors involve  auxiliary parameters, determining these parameters within the QCDSR framework requires careful consideration of several key points:
\\
1. Physical range stability: The physical range should be limited to ensure stability in variable changes.
2.  OPE convergence: The OPE series must converge.
3. Pole contribution dominance: The pole contribution (PC)  should dominate over continuum contributions.
\\
 In this section, these ranges are determined through numerical analysis and by considering the above criteria. Table \ \ref{inputParameter} presents the input data used for the analysis.  We choose the structures   $ \slashed {p}'\gamma_{\mu} \slashed{p} $,  $p_{\mu}\slashed {p'}\slashed {p}$,    $p'_{\mu}\slashed {p'}\slashed {p}$,    $\slashed {p}'\gamma_{\mu} \slashed{p}\gamma_5 $,  $p_{\mu}\slashed {p}'\slashed {p}\gamma_5$  and  $p'_{\mu}\slashed {p}'\slashed {p}\gamma_5$   for the form factors $ F_1 $,  $ F_2 $,  $F_3  $,  $G_1  $, $G_2  $ and $G_3  $,  respectively which lead to more stable and reliable results based on the standard prescriptions of the method that we are going to discuss. 
\begin{table}[h!]
\caption{Input parameters used in calculations.}\label{inputParameter}
\begin{tabular}{|c|c|}
\hline 
Parameters                                             &  Values  \\
\hline 
$m_d$                                                    &$4.70\pm0.07~ \mathrm{MeV}$\cite{ParticleDataGroup:2024cfk}\\
$m_s$                                                  & $93.5\pm0.8~ \mathrm{MeV}$\cite{ParticleDataGroup:2024cfk}\\
$ m_c$                                                 & $1.2730\pm0.0046~ \mathrm{GeV}$ \cite{ParticleDataGroup:2024cfk}\\
$ m_b$                                                 & $4.183\pm0.007~ \mathrm{GeV}$ \cite{ParticleDataGroup:2024cfk}\\
$ m_e $                                                & $ 0.51~~\mathrm{MeV}$\cite{ParticleDataGroup:2024cfk}\\
$ m_\mu $                                              & $ 0.1056~~\mathrm{GeV}$ \cite{ParticleDataGroup:2024cfk}\\
$ m_\tau $                                             & $ 1.776~~\mathrm{GeV}$ \cite{ParticleDataGroup:2024cfk}\\
$ m_{\Xi_b}$                                       & $ 5797.0\pm0.6 \mathrm{MeV}$ \cite{ParticleDataGroup:2024cfk}\\
$ m_{\Xi^{'}_{b}}$                                       & $ 5935.1\pm0.5\mathrm{MeV}$ \cite{ParticleDataGroup:2024cfk}\\
$ m_{\Xi_c} $                                      & $ 2470.44\pm0.28\mathrm{MeV}$  \cite{ParticleDataGroup:2024cfk} \\
$ m_{\Xi^{'}_{c}} $                                      & $ 2578.7\pm0.5\mathrm{MeV}$ \cite{ParticleDataGroup:2024cfk} \\
$ G_{F} $                                              & $ 1.17\times 10^{-5} \mathrm{GeV^{-2}}$\cite{ParticleDataGroup:2024cfk}\\
$ V_{cb} $                                             & $ 39\pm1.1\times 10^{-3}$  \cite{ParticleDataGroup:2024cfk}\\
$ m^2_0 $                                              & $ (0.8\pm0.2) \mathrm{GeV^2}$\cite{Belyaev:1982sa,Belyaev:1982cd,Ioffe:2005ym} \\
$\tau_{\Xi_b} $                                    & $ 1.572\pm0.040\times 10^{-12} s$ \cite{ParticleDataGroup:2024cfk}\\
$\langle \bar{u} u\rangle$         & $-(0.24\pm0.01)^3 \mathrm{GeV^3}$ \cite{Belyaev:1982sa,Belyaev:1982cd} \\
$\langle \bar{s} s\rangle$           & $(0.8\pm0.1) \langle \bar{u}u\rangle \mathrm{GeV^3}$ \cite{Belyaev:1982sa,Belyaev:1982cd} \\
$\langle0|\frac{1}{\pi}\alpha_s G^2|0\rangle$          &$ (0.012\pm0.004)\mathrm{GeV^4}$ \cite{Belyaev:1982sa,Ioffe:2005ym,Belyaev:1982cd}\\
$\lambda_{\Xi_b}$                               &$0.054\pm0.012  \mathrm{GeV^3}$\cite{Aliev:2018lcs}\\
$\lambda_{\Xi^{'}_{b}}$                               & $0.079\pm0.020  \mathrm{GeV^3}$ \cite{Wang:2009cr} \\
$\lambda_{\Xi_c}$                               &$0.036\pm0.021  \mathrm{GeV^3}$\cite{Aliev:2018ube}\\
$\lambda_{\Xi^{'}_{c}}$                               & $0.040\pm0.005  \mathrm{GeV^3}$\cite{Agaev:2020fut}\\
\hline
\end{tabular}
\end{table}

\begin{figure}
\includegraphics[totalheight=5.5cm,width=5.9cm]{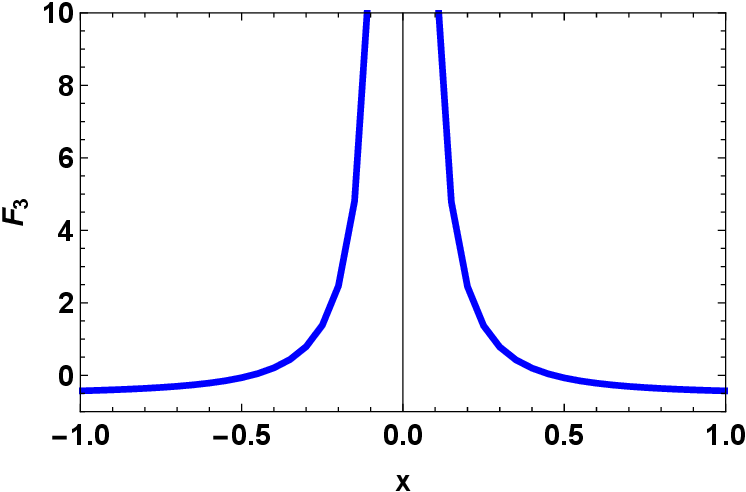}
\includegraphics[totalheight=5.5cm,width=5.9cm]{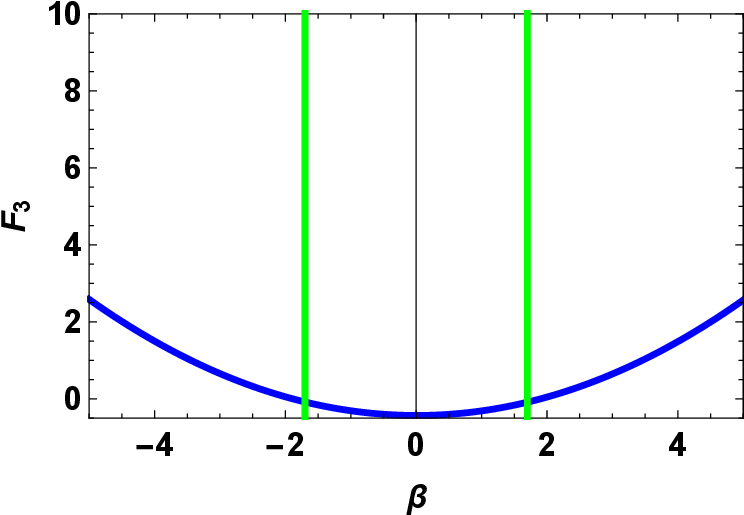}
\caption{The form factor $F_3 $ as a function of $x$ and $\beta$ at $q^2=0$ and average values of other auxiliary parameters.}\label{Fx}
\end{figure}

The first mathematical parameter introduced in the calculations is $\beta$, the mixing current parameter, which can has an unlimited range from negative infinity to positive infinity. To define its physical range, it is reparametrized as:
\begin{equation} \label{x}
x=cos\theta ~~ \mbox{where}~~ \beta=tan^{-1}\theta,
\end{equation}
with $x$ contained to the region $[-1,1]$. A stable range for the form factors is selected where they remain approximately constant as $x$ changes. Based on numerical analysis, the ranges $-1\leq x \leq -0.5$ and $0.5\leq x\leq 1$ are chosen.   As an example,  in  Fig. \ref{Fx},  left panel,  we show how the form factor $ F_3 $  demonstrates good stability with respect to the variations of $ x $ in the chosen region. In the right panel of this figure we also demonstrate the variations of $ F_3 $ with respect to $  \beta$. The vertical lines show the borders of the selected regions.

The second set of auxiliary parameters are the continuum thresholds ($s_0$ and $s'_0$), which are incorporated into the calculations through the application of quark-hadron duality. This involves modifying the upper limits to reduce contributions from higher states. The thresholds are determined by considering the energy of the first excited state with the same quantum numbers as the interpolating currents  at each channel. The chosen ranges are:
\begin{eqnarray}
&&(m_{\Xi_b}+0.1)^2~ \mathrm{GeV^2} \leq s_{0} \leq (m_{\Xi_b}+0.5)^2~ \mathrm{GeV^2},\notag\\
\mbox{and} \notag\\
&&(m_{\Xi_c}+0.1)^2~\mathrm{GeV^2}\leq s'_{0} \leq (m_{\Xi_c}+0.5)^2~ \mathrm{GeV^2}.
\end{eqnarray}
The last auxiliary parameter to be fixed are the Borel parameters  $M^2$ and $M'^2$. To establish their upper limits, the dominance condition of the PC over higher states and continuum is used:

\begin{equation} \label{PC}
PC=\frac{\Pi(M^2,M'^2,s_0,s'_0)}{\Pi(M^2,M'^2,{\infty},{\infty})}\geq\frac{1}{2}.
\end{equation}
The lower limits are set so that higher dimensional non-perturbative operators contribute minimally, ensuring OPE series convergence:

\begin{equation} \label{PC2}
R(M^2_{min},M'^2_{min})=\frac{\Pi^{dim6}(M^2_{min},M'^2_{min},s_0,s'_0)}{\Pi(M^2_{min},M'^2_{min},s_0,s'_0)}\leq0.05.
\end{equation}
Thus,  the Borel parameters are constrained within the following ranges:

\begin{eqnarray}
 &&8~\mathrm{GeV^2}\leq M^2 \leq 10~\mathrm{GeV^2}, \notag\\
\mbox{and} \notag\\
&&6~\mathrm{GeV^2} \leq M'^2 \leq 8~\mathrm{GeV^2}.
\end{eqnarray}

\begin{figure}[h!] 
\includegraphics[totalheight=5.8cm,width=5.9cm]{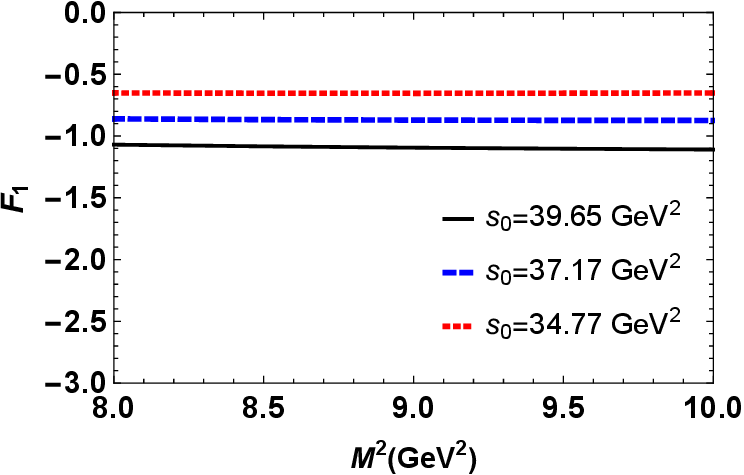}
\includegraphics[totalheight=5.8cm,width=5.9cm]{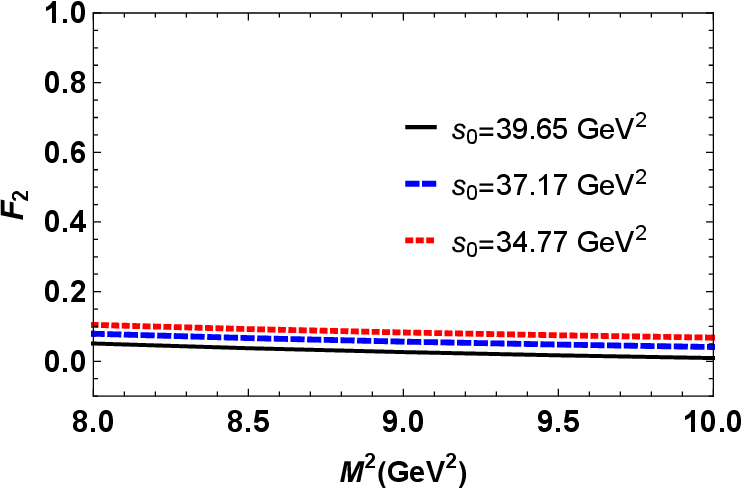}
\includegraphics[totalheight=5.8cm,width=5.9cm]{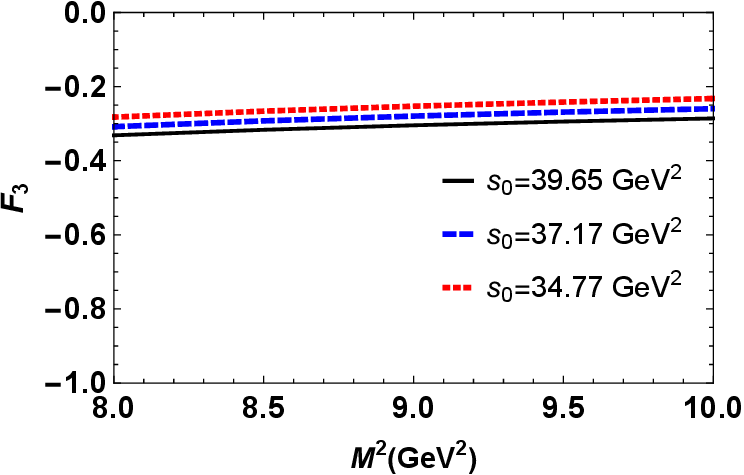}
\includegraphics[totalheight=5.8cm,width=5.9cm]{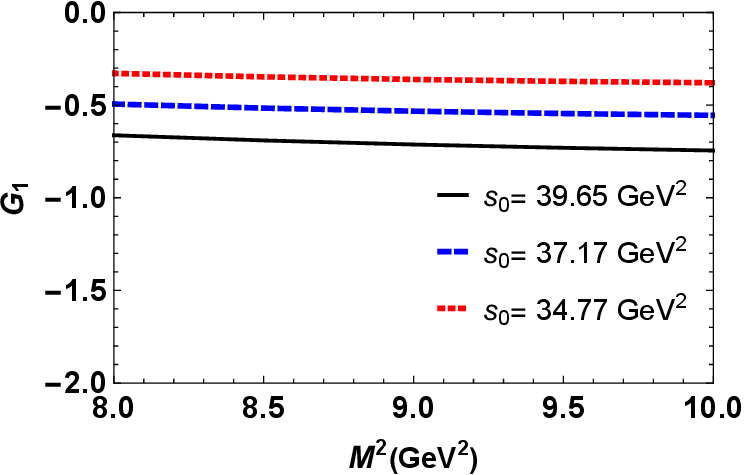}
\includegraphics[totalheight=5.8cm,width=5.9cm]{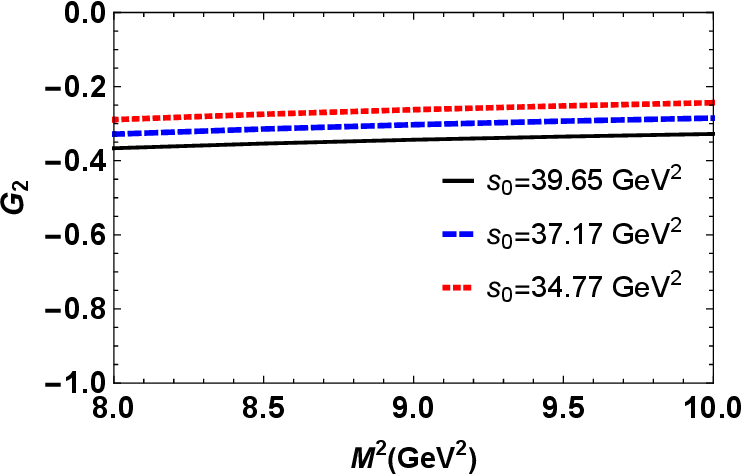}
\includegraphics[totalheight=5.8cm,width=5.9cm]{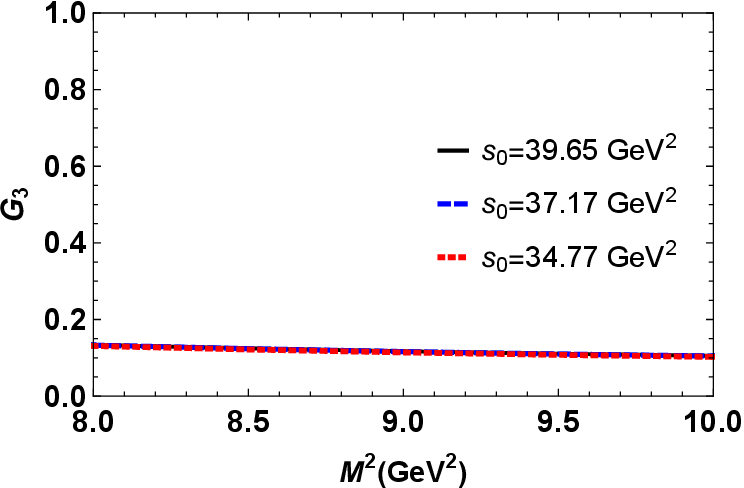}
\caption{Form factors as functions of the Borel parameter $M^2$ at  various values 
of the parameter $s_0$,   $q^2=0$ and average values of other auxiliary parameters.}\label{Fig:BorelMs}
\end{figure}
\begin{figure}[h!]
\includegraphics[totalheight=5.8cm,width=5.9cm]{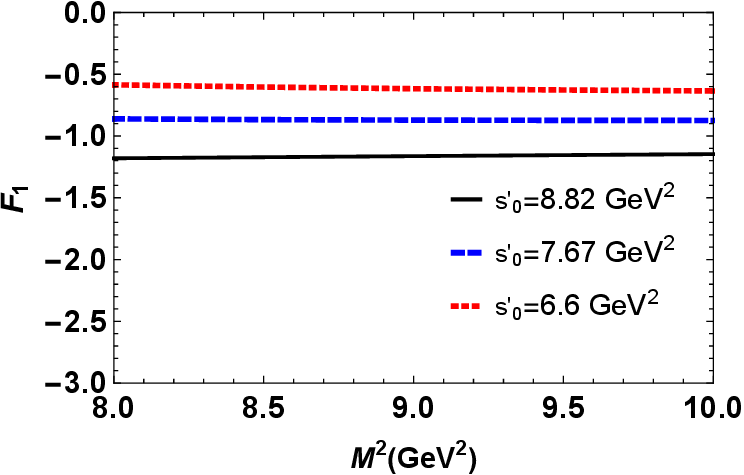}
\includegraphics[totalheight=5.8cm,width=5.9cm]{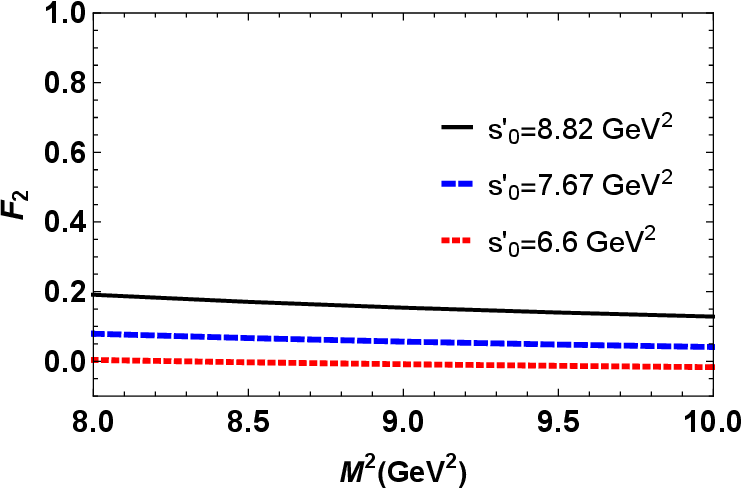}
\includegraphics[totalheight=5.8cm,width=5.9cm]{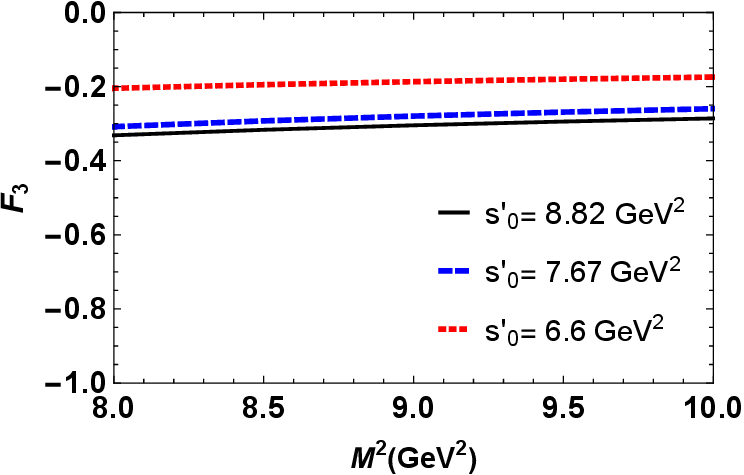}
\includegraphics[totalheight=5.8cm,width=5.9cm]{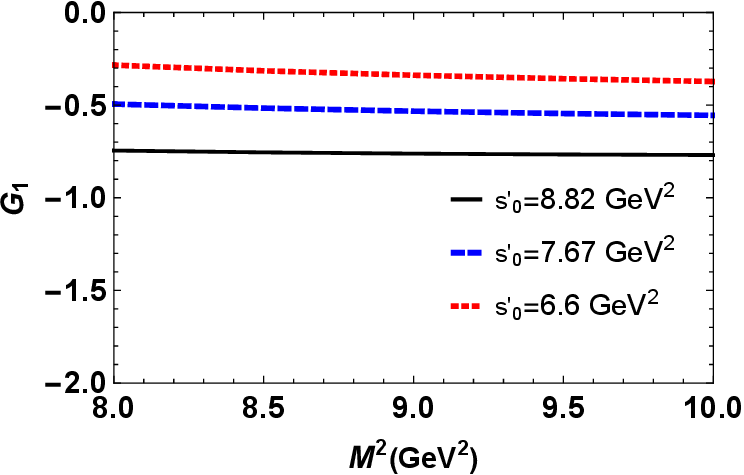}
\includegraphics[totalheight=5.8cm,width=5.9cm]{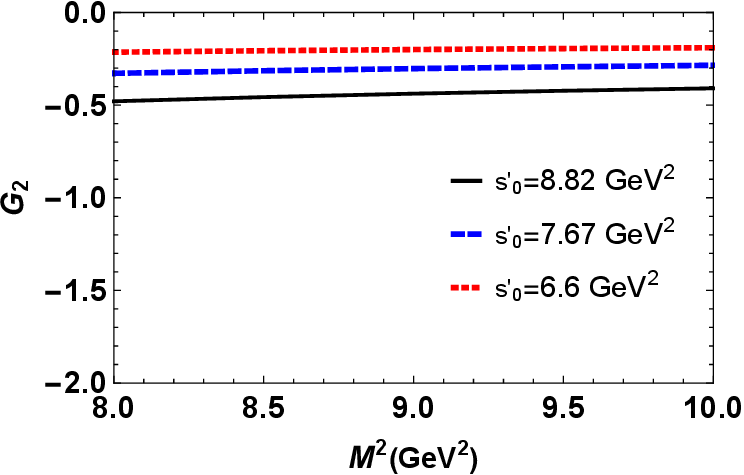}
\includegraphics[totalheight=5.8cm,width=5.9cm]{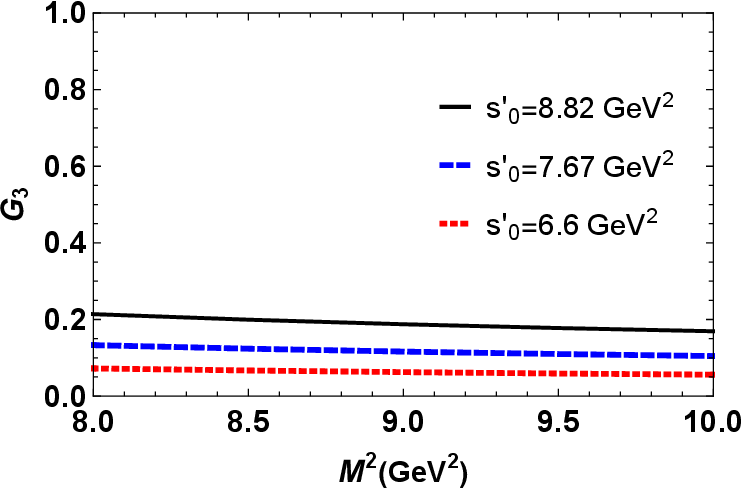}
\caption{Form factors as functions of the Borel parameter $M^2$ at various values 
of the parameter $s'_0$,  $q^2=0$ and average values of other auxiliary parameters.} \label{Fig:BorelM's}
\end{figure}

\begin{figure}[h!] 
\includegraphics[totalheight=5.8cm,width=5.9cm]{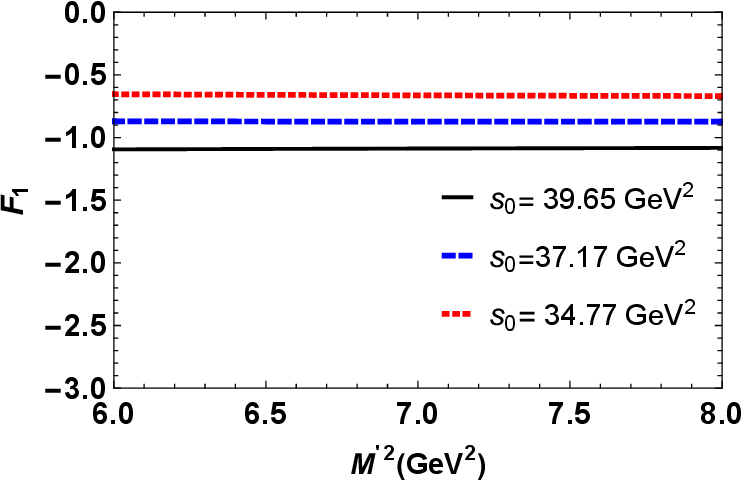}
\includegraphics[totalheight=5.8cm,width=5.9cm]{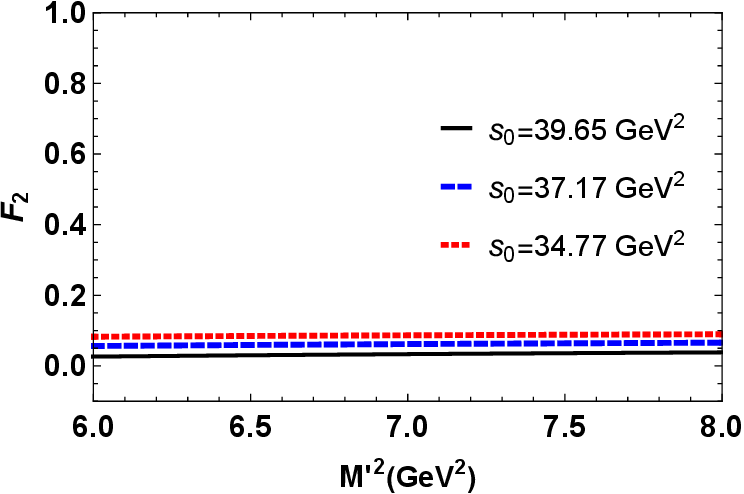}
\includegraphics[totalheight=5.8cm,width=5.9cm]{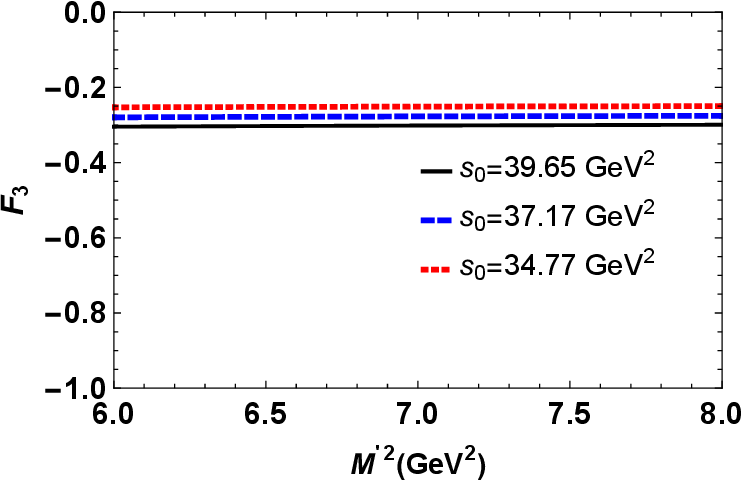}
\includegraphics[totalheight=5.8cm,width=5.9cm]{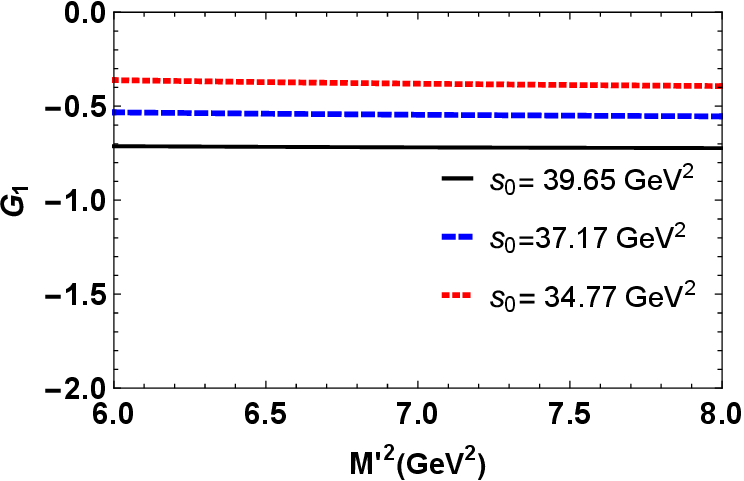}
\includegraphics[totalheight=5.8cm,width=5.9cm]{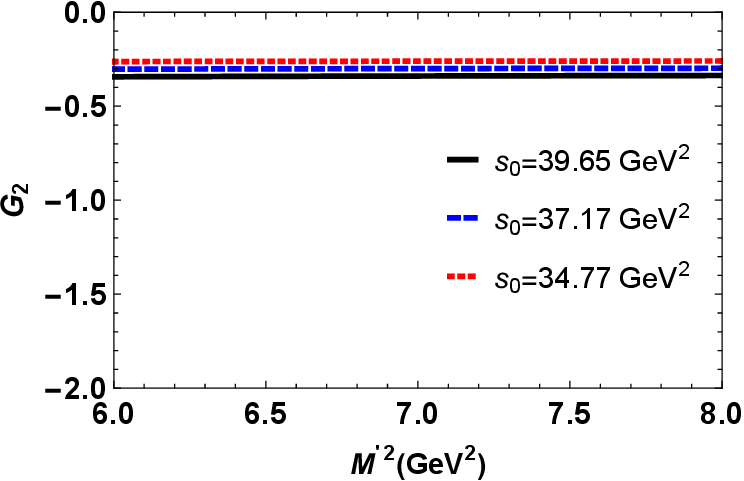}
\includegraphics[totalheight=5.8cm,width=5.9cm]{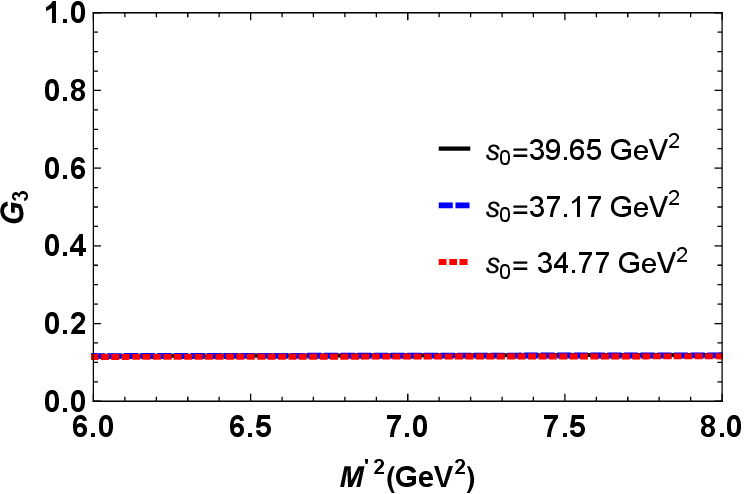}
\caption{Form factors  as functions of the Borel parameter $M'^2$ at  various values 
of the parameter $s_0$,   $q^2=0$ and average values of other auxiliary parameters.}\label{Fig:BorelMs'}
\end{figure}
\begin{figure}[h!]
\includegraphics[totalheight=5.9cm,width=5.9cm]{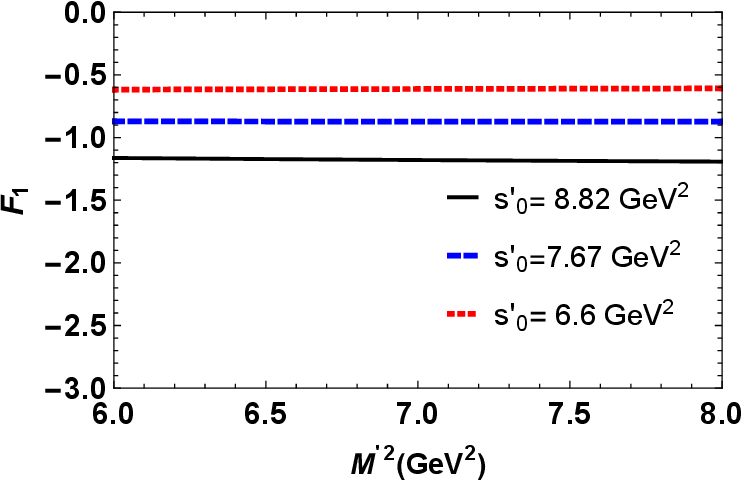}
\includegraphics[totalheight=5.9cm,width=5.9cm]{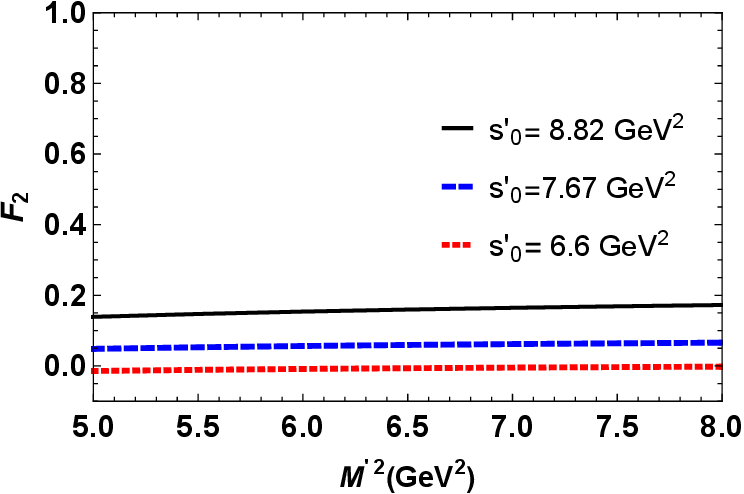}
\includegraphics[totalheight=5.9cm,width=5.9cm]{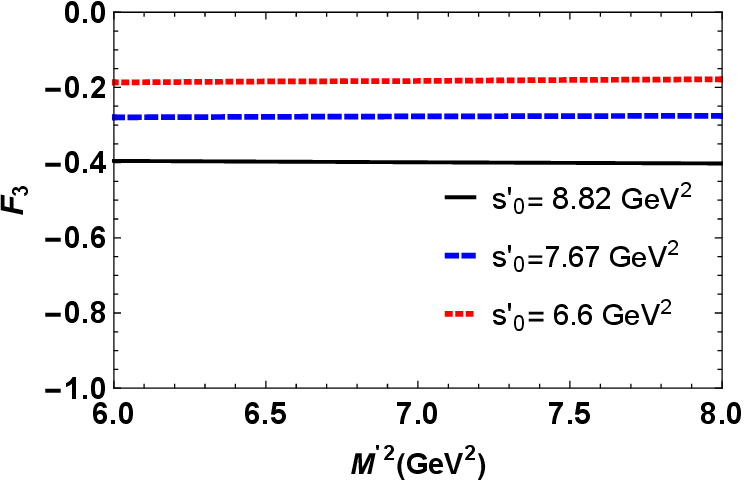}
\includegraphics[totalheight=5.9cm,width=5.9cm]{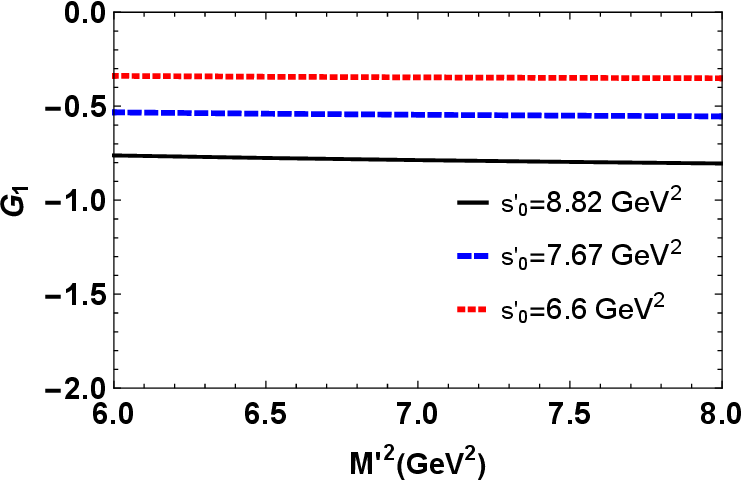}
\includegraphics[totalheight=5.9cm,width=5.9cm]{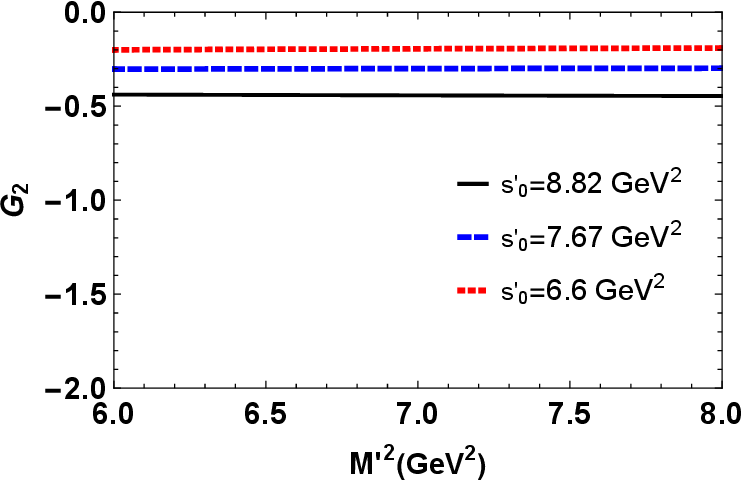}
\includegraphics[totalheight=5.9cm,width=5.9cm]{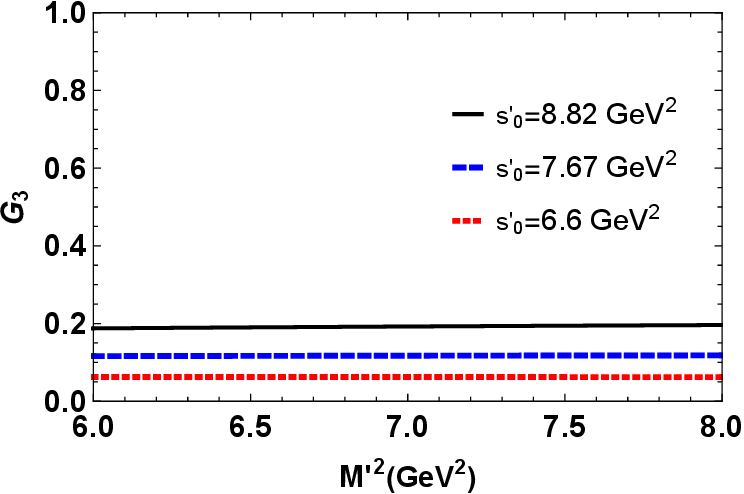}
\caption{Form factors as functions of the Borel parameter $M'^2$ at various values 
of the parameter $s'_0$,  $q^2=0$ and average values of other auxiliary parameters.} \label{Fig:BorelM's'}
\end{figure}

\begin{figure}[h!] 
\includegraphics[totalheight=5cm,width=5.8cm]{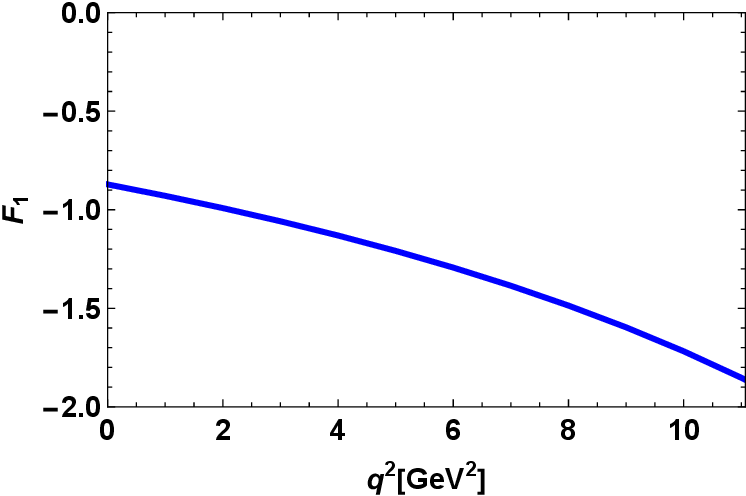}
\includegraphics[totalheight=5cm,width=5.8cm]{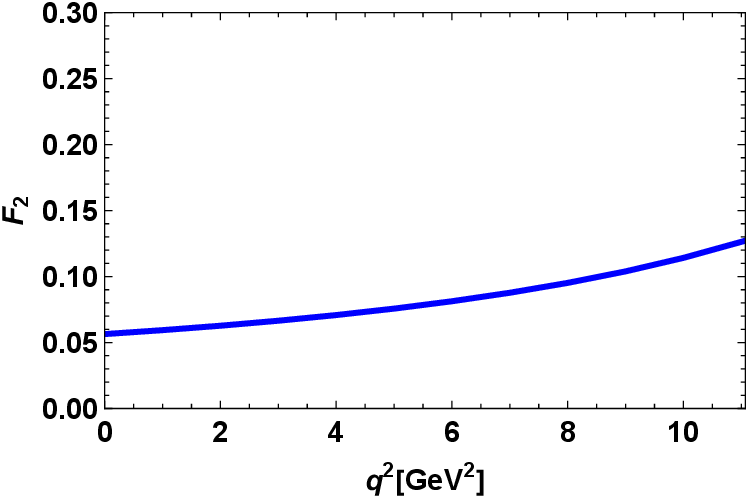}
\includegraphics[totalheight=5cm,width=5.8cm]{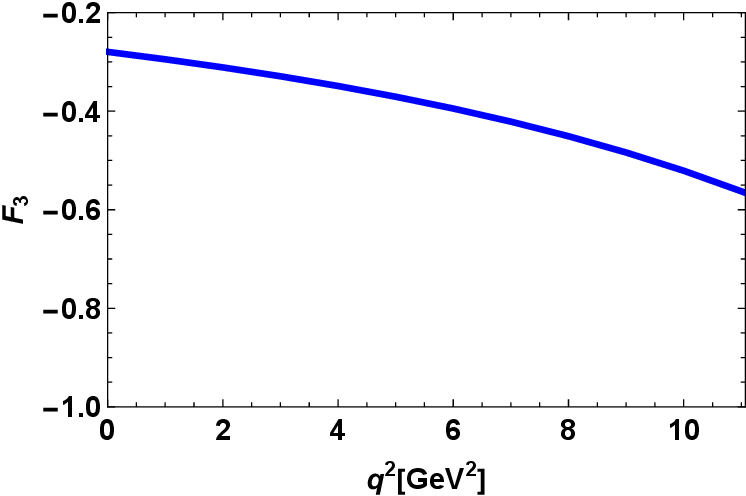}
\includegraphics[totalheight=5cm,width=5.8cm]{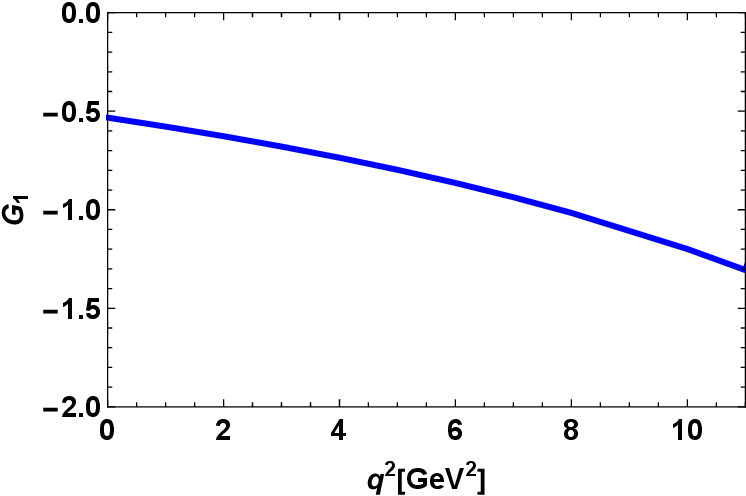}
\includegraphics[totalheight=5cm,width=5.8cm]{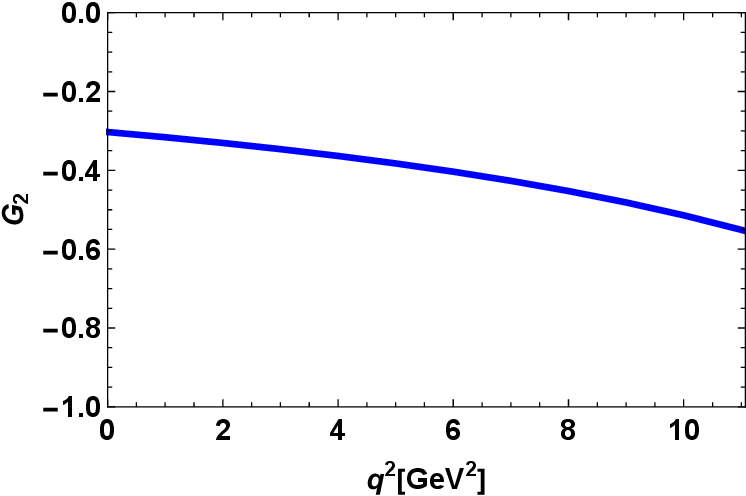}
\includegraphics[totalheight=5cm,width=5.8cm]{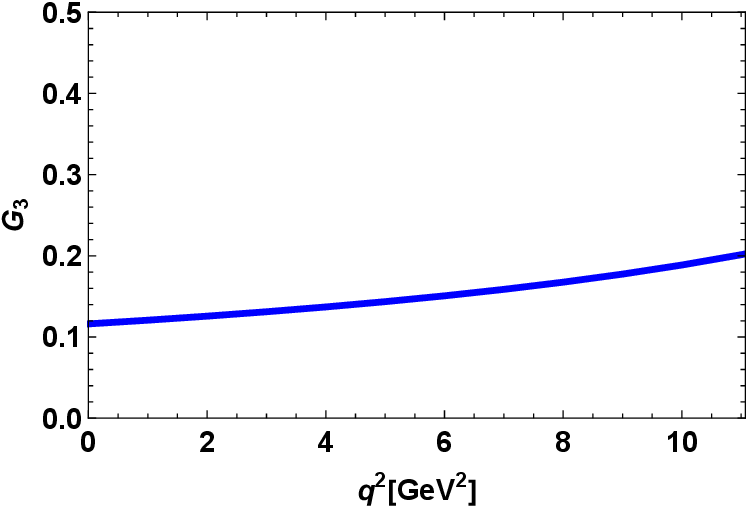}
\caption{The form factors $F_1$,  $F_2$, $F_3$,  $G_1$ , $G_2$ and $G_3$ as functions of $q^2$ at average values of auxiliary parameters.}\label{Fig:formfactor1}
\end{figure}

The stability of the form factors with respect to parameter changes, as observed in Figs.  \ref{Fig:BorelMs}, \ref{Fig:BorelM's}, \ref{Fig:BorelMs'} and \ref{Fig:BorelM's'} support the chosen physical ranges for the auxiliary parameters.
After determining the appropriate ranges of the auxiliary parameters, we plot the form factors as a function of $q^2$.  The weak decay form factors exhibit a behavior where they move away from the coordinate axis as $q^2$ increases, consistent with the expectations for weak decays (Fig.  \ref{Fig:formfactor1}).  Analyzing the form factors as a function of $q^2$ allows us to obtain a fitting function for the form factors:
\begin{equation} \label{fitffunction}
{\cal F}(q^2)=\frac{{\cal
F}(0)}{\displaystyle\left(1-a_1\frac{q^2}{m^2_{\Xi_b}}+a_2
\frac{q^4}{m_{\Xi_b}^4}+a_3\frac{q^6}{m_{\Xi_b}^6}+a_4\frac{q^8}{m_{\Xi_b}^8}\right)}.
\end{equation}
This function has been selected based on the behavior of the weak decay form factors as function of  $q^2$ in order to accurately model their curved and non-linear variations. This choice aims for a better fit to the sum rules results and a more precise description of the increasing trend of the form factors as $q^2$ increases.
The values of the parameters,  ${\cal F}(0)$,  $ a_1$,  $a_2$,  $a_3$ and $a_4$, are obtained by using the average over suitable ranges of  auxiliary parameters such as $M^2$, $M'^2$, $s_0$, and $s'_0$  for $\Xi_{b}\rightarrow
\Xi_{c}{\ell}\bar\nu_{\ell}$ transition. The results are presented  in Table \ref{Tab:parameterfit1}.

\begin{table}[h!]
\caption{Parameters of the fit functions for different form factors corresponding to $\Xi_b\to\Xi_c l\bar\nu_{\ell}$ transition.}\label{Tab:parameterfit1}
\begin{ruledtabular}
\begin{tabular}{|c|c|c|c|c|c|c|}
            & $F_1(q^2) $ & $F_2(q^2)$  & $F_3(q^2)$   & $G_1(q^2) $ & $G_2(q^2)$  & $G_3(q^2) $       \\
\hline
${\cal F}(q^2=0)$ & $-0.87\pm0.21$        & $0.06\pm0.01$      & $-0.28\pm0.08$     & $-0.53\pm0.11$  & $-0.30\pm0.09$  & $0.12\pm0.05$  \\
$a_1$           & $2.11$          & $1.68$            & $1.73$           & $2.68$           & $1.41$            &$ 1.27$              \\
$a_2$           & $1.20$          & $-0.29$            & $0.54$             & $3.89$          & $0.08$           & $-0.06$           \\
$a_3$           & $-1.60$         & $0.48$            & $0.18$          & $-4.48$          & $0.05$           & $-0.09$           \\
$a_4$           & $0.78$           & $1.05$           & $-0.005$            & $2.74$         & $0.22$          & $0.26$           \\
\end{tabular}
\end{ruledtabular}
\end{table}
\begin{table}[h!]
\caption{Parameters of the fit functions for different form factors corresponding to $\Xi_b\to\Xi^{'}_c l\bar\nu_{\ell}$ transition.}\label{Tab:parameterfit2}
\begin{ruledtabular}
\begin{tabular}{|c|c|c|c|c|c|c|}
            & $F_1(q^2) $ & $F_2(q^2)$  & $F_3(q^2)$   & $G_1(q^2) $ & $G_2(q^2)$  & $G_3(q^2) $       \\
\hline
${\cal F}(q^2=0)$ & $-0.55\pm0.14$        & $-0.73\pm0.12$      & $0.065\pm0.014$     & $0.015\pm0.009$  & $-0.79\pm0.14$  & $0.069\pm0.013$  \\
$a_1$           & $1.33$          & $1.29$            & $2.05$           & $12.95$           & $1.28$            &$ 2.06$              \\
$a_2$           & $0.25$          & $0.24$            & $1.24$             & $80.31$          & $0.23$           & $1.26$           \\
$a_3$           & $0.07$         & $0.06$            & $-0.16$          & $-235.05$          & $0.06$           & $-0.17$           \\
$a_4$           & $0.01$           & $0.005$           & $-0.037$            & $260.17$         & $0.007$          & $-0.035$           \\
\end{tabular}
\end{ruledtabular}
\end{table}

\begin{table}[h!]
\caption{Parameters of the fit functions for different form factors corresponding to $\Xi^{'}_b\to\Xi^{'}_c l\bar\nu_{\ell}$ transition.}\label{Tab:parameterfit3}
\begin{ruledtabular}
\begin{tabular}{|c|c|c|c|c|c|c|}
            & $F_1(q^2) $ & $F_2(q^2)$  & $F_3(q^2)$   & $G_1(q^2) $ & $G_2(q^2)$  & $G_3(q^2) $       \\
\hline
${\cal F}(q^2=0)$ & $0.44\pm0.13$     &  $-0.45\pm0.10$      & $-0.18\pm0.06$     & $0.028\pm0.015$  & $-0.75\pm0.18$  & $0.15\pm0.04$  \\
$a_1$           & $1.49$          & $1.24$            & $1.66$           & $1.31$           & $1.36$           &$ 1.55$              \\
$a_2$           & $0.46$          & $0.28$            & $0.38$           & $1.00$          & $0.17$           & $0.23$           \\
$a_3$           & $-0.05$         & $0.05$            & $0.18$          & $-0.71$          & $0.06$         & $0.14$           \\
$a_4$           & $-0.003$       & $0.23$           & $0.06$            & $0.33$         & $0.02$          & $0.13$           \\
\end{tabular}
\end{ruledtabular}
\end{table}

\begin{table}[h!]
\caption{Parameters of the fit functions for different form factors corresponding to $\Xi^{'}_b\to\Xi_c l\bar\nu_{\ell}$ transition.}\label{Tab:parameterfit4}
\begin{ruledtabular}
\begin{tabular}{|c|c|c|c|c|c|c|}
            & $F_1(q^2) $ & $F_2(q^2)$  & $F_3(q^2)$   & $G_1(q^2) $ & $G_2(q^2)$  & $G_3(q^2) $       \\
\hline
${\cal F}(q^2=0)$ & $-0.17\pm0.04$        & $-0.21\pm0.05$      & $0.015\pm0.005$     & $0.031\pm0.006$  & $-0.21\pm0.06$  & $0.031\pm0.008$  \\
$a_1$           & $1.37$          & $1.30$            & $2.14$           & $1.07$           & $1.33$            &$ 1.79$              \\
$a_2$           & $0.26$          & $0.22$            & $1.34$           & $0.07$          & $0.24$           & $0.75$           \\
$a_3$           & $0.075$        & $0.06$            & $-0.17$          & $0.02$          & $0.07$           & $0.06$           \\
$a_4$           & $0.015$        & $0.008$          & $-0.04$            & $0.01$         & $0.01$          & $-0.03$           \\
\end{tabular}
\end{ruledtabular}
\end{table}

\begin{figure}[h!] 
\includegraphics[totalheight=5cm,width=5.8cm]{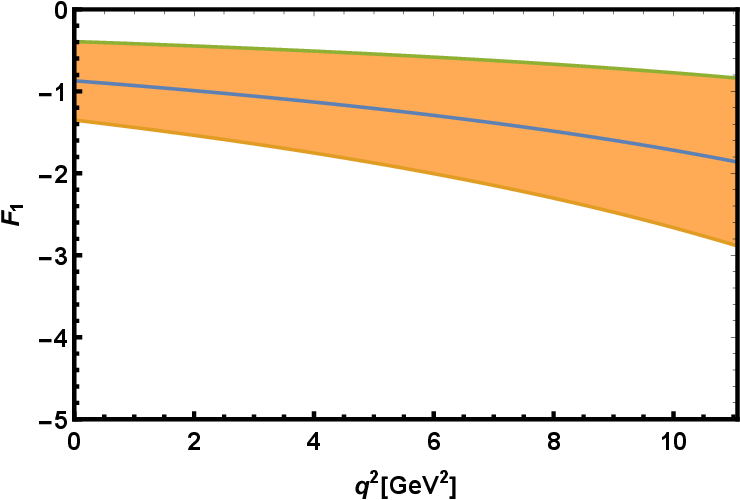}
\includegraphics[totalheight=5cm,width=5.8cm]{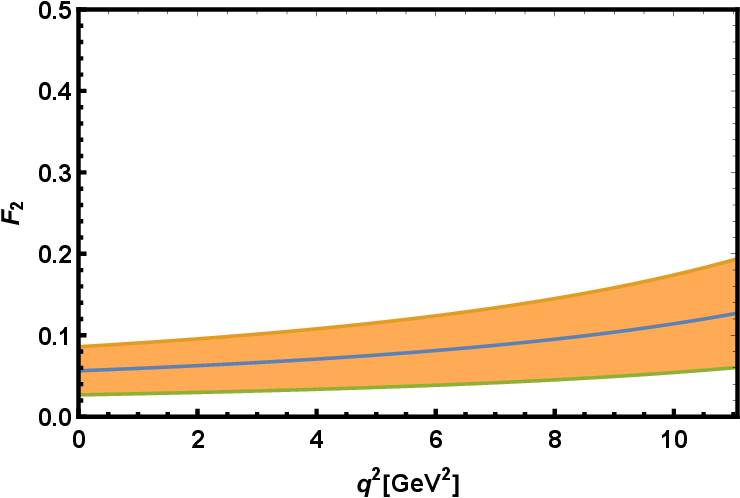}
\includegraphics[totalheight=5cm,width=5.8cm]{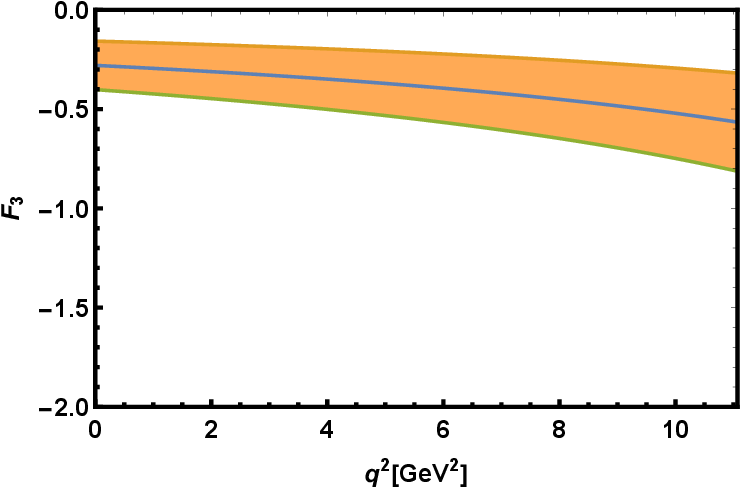}
\includegraphics[totalheight=5cm,width=5.8cm]{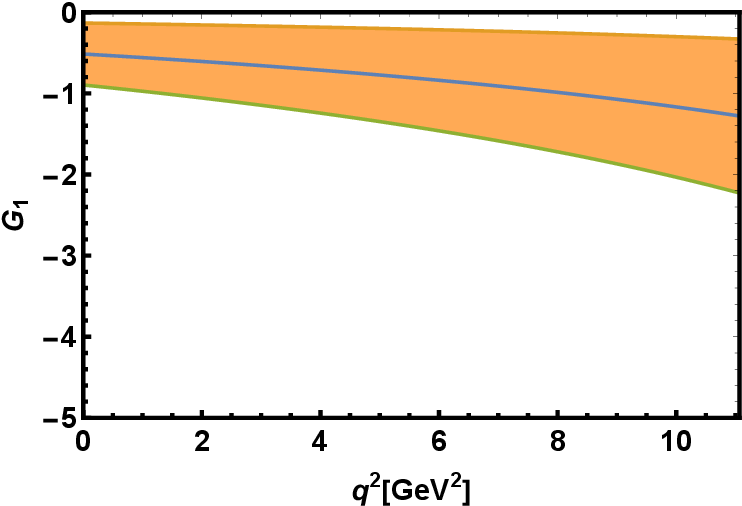}
\includegraphics[totalheight=5cm,width=5.8cm]{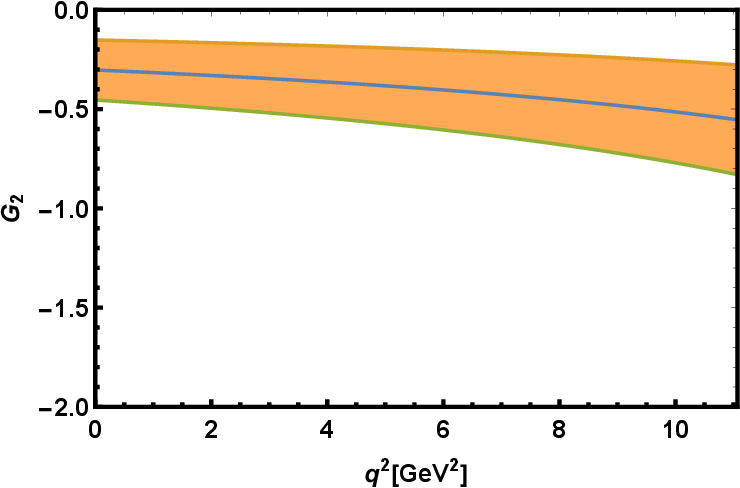}
\includegraphics[totalheight=5cm,width=5.8cm]{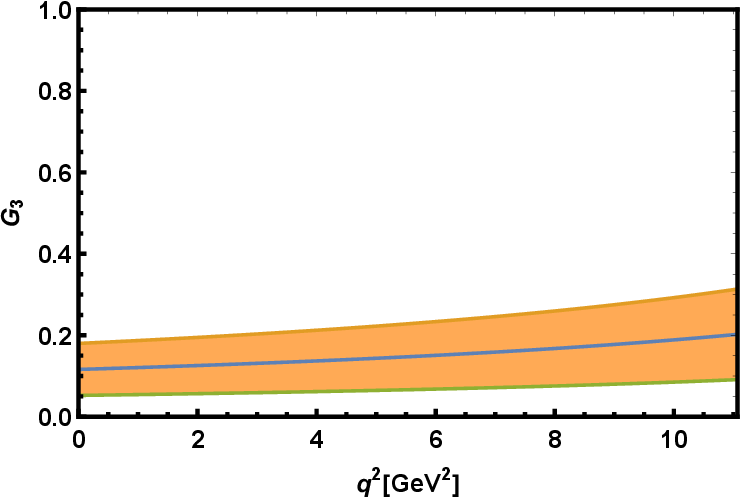}
\caption{The  fit functions of form factors with their uncertainties  as functions of $q^2$.}\label{Fig:formfactorserror1}
\end{figure}
The uncertainties presented in the Table \ref{Tab:parameterfit1} are  associated with the uncertainties in determining the working ranges of  the auxiliary parameters. In Fig. \ref {Fig:formfactorserror1} the form factors are plotted considering  these uncertainties. 
We repeat all the analyses to find the fitting parameters of other decay modes considered in the present study.  The results are depicted in Tables \ref{Tab:parameterfit2},  \ref{Tab:parameterfit3} and \ref{Tab:parameterfit4}.
 In the next step, we obtain the decay rates and branching ratios within the allowed range $0\leq q^2 \leq (m_{\Xi_b}-m_{\Xi_c})^2$ using the fitting functions of the form factors.

\section{Decay Width and Branching Ratio}~\label{Sec4}
The main objective of this work is to calculate the decay rates and branching ratios of  $\Xi^{(')}_{b}\rightarrow
\Xi^{(')}_{c}{\ell}\bar\nu_{\ell}$ decays. After obtaining the fit functions for the form factors, we can determine the decay rates using the differential decay width formula \cite{Faustov:2016pal,Gutsche:2014zna,Migura:2006en,Korner:1994nh,Bialas:1992ny}: 

\begin{equation}\label{eq:dgamma}
\frac{d\Gamma(\Xi_b\to\Xi_c\ell\bar\nu_\ell)}{dq^2}=\frac{G_F^2}{(2\pi)^3}
|V_{cb}|^2\frac{\lambda^{1/2}(q^2-m_\ell^2)^2}{48M_{\Xi_b}^3q^2}{\cal
H}_{tot}(q^2),
\end{equation}

where $\lambda\equiv\lambda(m^2_{\Xi_b}, m^2_{\Xi_c}, q^2)=m^4_{\Xi_b}+m^4_{\Xi_c}+q^4-2(m^2_{\Xi_b}m^2_{\Xi_c}+m^2_{\Xi_b}q^2+m^2_{\Xi_c}q^2)$, $m_l$ is the lepton mass and ${\cal H}_{tot}(q^2)$ represents the total helicity amplitude.
The total helicity amplitude is composed of:

\begin{equation}
 \label{eq:hh}
 {\cal H}_{tot}(q^2)=[{\cal H}_U(q^2)+{\cal H}_L(q^2)] \left(1+\frac{m_\ell^2}{2q^2}\right)+\frac{3m_\ell^2}{2q^2}{\cal H}_S(q^2) ,
\end{equation}
where the following components for relevant parity-conserving helicity structures are entered:
\begin{eqnarray}
  \label{eq:hhc}
&&{\cal H}_U(q^2)=|H_{+1/2,+1}|^2+|H_{-1/2,-1}|^2,\notag\\
&&{\cal H}_L(q^2)=|H_{+1/2,0}|^2+|H_{-1/2,0}|^2,\notag\\
&&{\cal H}_S(q^2)=|H_{+1/2,t}|^2+|H_{-1/2,t}|^2.\notag\\
\end{eqnarray}
 These helicity amplitudes are derived using the baryon form factors  $F_i$ and $G_i$:

\begin{eqnarray}
  \label{eq:ha}
  H^{V,A}_{+1/2,\, 0}&=&\frac1{\sqrt{q^2}}{\sqrt{2m_{\Xi_b}m_{\Xi_c}(\alpha\mp 1)}}
[(m_{\Xi_b} \pm m_{\Xi_c}){\cal F}^{V,A}_1(\alpha) \pm m_{\Xi_c}
(\alpha\pm 1){\cal F}^{V,A}_2(\alpha)\cr
&& \pm m_{\Xi_{b}} (\alpha\pm 1){\cal F}^{V,A}_3(\alpha)],\cr
 H^{V,A}_{+1/2,\, 1}&=&-2\sqrt{m_{\Xi_b}m_{\Xi_c}(\alpha\mp 1)}
 {\cal F}^{V,A}_1(\alpha),\cr
H^{V,A}_{+1/2,\, t}&=&\frac1{\sqrt{q^2}}{\sqrt{2m_{\Xi_b}m_{\Xi_c}(\alpha\pm 1)}}
[(m_{\Xi_b} \mp m_{\Xi_c}){\cal F}^{V,A}_1(\alpha) \pm(m_{\Xi_b}- m_{\Xi_c}\alpha
){\cal F}^{V,A}_2(\alpha)\cr
&& \pm (m_{\Xi_{b}} \alpha- m_{\Xi_c}){\cal F}^{V,A}_3(\alpha)],
\end{eqnarray}

where 
$$\alpha=\frac{m_{\Xi_b}^2+m_{\Xi_{c}}^2-q^2}
{2m_{\Xi_b}m_{\Xi_{c}}}.$$ 
To write the formula more neatly, ${\cal F}$ has been used, which represents the vector form factors (${\cal F}^V_i\equiv F_i$) and axial vector ( ${\cal F}^A_i\equiv G_i$) types, the upper signs refer to vector currents, while the lower signs refer to axial vectors ones.  $H^{V,A}_{h',\,h_W}$ correspond to  the helicity amplitudes for weak decays including the vector (V) and the axial vector (A) currents and  $(h',h_W)$ indexes indicate the helicities of the final baryon and the virtual W-boson. For the negative amplitude of the helicity, the  following relation holds:
\begin{eqnarray}
 H^{V,A}_{-h',\,-h_W}=\pm H^{V,A}_{h',\,h_W}.
\end{eqnarray}
The total amplitude for vector and axial vector currents is given by: 
\begin{eqnarray}
H_{h',\,h_W}=H^{V}_{h',\,h_W}-H^{A}_{h',\,h_W}.
\end{eqnarray}

Using the fit functions of the form factors from Table \ref {Tab:parameterfit1}, we evaluate the decay rates and  branching fractions for different lepton channels in $\Xi_{b}\rightarrow
\Xi_{c}{\ell}\bar\nu_{\ell}$  transition.  We present and compare our results  with other existing predictions in Tables \ref{DECAY} and \ref{BR}. The results obtained from our calculations,   in the Tables \ref{DECAY} and \ref{BR},  are consistent with predictions of other studies, taking into account the reported uncertainties.

The calculation of the ratio of branching fractions  in $\tau$ and $e$/$\mu$ channels is useful for comparing the SM predictions with future experimental results, which we find here:
\begin{eqnarray}
R_{\Xi_c}=\frac{Br[\Xi_b\rightarrow \Xi_c\tau
\overline{\nu}_\tau]}{Br[\Xi_b\rightarrow \Xi_c
(e,\mu)\overline{\nu}_{(e,\mu)}]}=0.34\pm0.15
\end{eqnarray} 
Our result is consistent with that of  Ref.  \cite{Faustov:2018ahb} with  $R_{\Xi_c}=0.325\pm0.010$ for the decay channel  $\Xi_{b}\rightarrow
\Xi_{c}{\ell}\bar\nu_{\ell}$ .
\begin{table}[h!]
\caption{Decay widths (in $\times10^{14}\mathrm{GeV}  $) for the semileptonic $\Xi_b\to\Xi_c \ell {\overline{\nu}}_\ell$ transition at different lepton  channels.}\label{DECAY}
\begin{ruledtabular}
\begin{tabular}{|c|c|c|}
                 &$\Gamma [\Xi_b\to\Xi_c (e,\mu) {\overline{\nu}}_{(e,\mu)}]$& $ \Gamma[\Xi_b\to \Xi_c \tau \overline{\nu}_{\tau}]$ \\
\hline
Present Work&  $3.43^{+1.83}_{-1.40}$ & $1.18^{+0.63}_{-0.48}$\\
\hline
The relativistic equasipotential approach\cite{Faustov:2018ahb}&$2.57$&$0.836$\\
\hline
The ralativistic quark model\cite{Ebert:2005ip}&$3.48$&-\\
\hline
The relativistic quasipotential equationl\cite{Ebert:2006rp} &$3.48$&-\\
\hline
The ralativistic three quark model\cite{Ivanov:1999pz} &$4.39$&-\\
\hline
The ralativistic three quark model\cite{Ivanov:1996fj}&$3.47$&-\\
\hline
The nonrelativistic constituent quark model\cite{Albertus:2004wj} &$3.28$&-\\ 
\hline
The nonrelativistic quark model\cite{Cheng:1995fe}&$3.49$&-\\
\hline
Bethe-Salpeter equation\cite{Ivanov:1998ya}&$4.22$&-\\
\hline
Covariant Quasipotential Approach\cite{Rusetsky:1997id}&$4.21$&-\\
\hline
QCDSR\cite{Zhao:2020mod}  & $3.80\pm0.33$&-\\
\hline
Light front \cite{Cardarelli:1998tq} &$3.74\pm1.5$&-\\
\hline
Light front \cite{Zhao:2018zcb} & $3.97$&-\\
\hline
Light front\cite{Zhao:2022vfr} &$3.97$&-\\
\hline
HQET\cite{Korner:1994nh}& $3.43$ &-\\
\hline
The spectator quark model\cite{Singleton:1990ye}&$ 4.74$&-\\
\hline
\end{tabular}
\end{ruledtabular}
\end{table}

\begin{table}[h!]
\caption{Branching ratios (in \%) of the semileptonic $\Xi_b\to\Xi_c \ell {\overline{\nu}}_\ell$ transition at different channels.}\label{BR}
\begin{ruledtabular}
\begin{tabular}{|c|c|c|c|c|c|c|c|}
                 & Present Work &The relativistic\cite{Faustov:2018ahb}&The relativistic \cite{Dutta:2018zqp}&Light front\cite{Cardarelli:1998tq}&Light front\cite{Zhao:2022vfr} &QCDSR\cite{Zhao:2020mod}   \\
\hline
$Br~[\Xi_b\to\Xi_c (e,\mu) {\overline{\nu}}_{(e,\mu)}]$ &$8.18^{+4.36}_{-3.34}$& $6.15$  &  $9.22$ &   $7.9\pm2.0$ & $9.42$&  $9.02\pm0.79$ \\
\hline
 $Br~[\Xi_b\to \Xi_c \tau \overline{\nu}_{\tau}]$  &$2.81^{+1.50}_{-1.15}$ &$2.00$ &$2.35$&-&-&- \\       
\end{tabular}
\end{ruledtabular}
\end{table}

We repeat all the above computational steps  and calculate the relevant decay rates,  branching  ratios and ratio of the branching fractions at different lepton channels for other transitions, i.e.,  $\Xi_b\to\Xi^{'}_c  \ell {\overline{\nu}}_\ell$,  $ \Xi^{'}_b\to\Xi^{'}_c   \ell {\overline{\nu}}_\ell$ and $  \Xi^{'}_b\to\Xi_c   \ell {\overline{\nu}}_\ell$ in  Tables \ref{DECAY2},  \ref{DECAY3} and \ref{DECAY4},  respectively.   No prior studies have investigated these decay channels to have  a direct comparison with our results. Only  Ref. \cite{Ebert:2006rp} has reported the width in one channel $\Gamma [\Xi^{'}_b\to\Xi^{'}_c e {\overline{\nu}}_{e}]=8.82 \times10^{15}\mathrm{GeV}$ which is very close to our related prediction. Note that,  we have the reported  lifetime of the $ \Xi_b $ in PDG allows us to find the branching ratio and ratios of the branching fractions at different channels. However, this is not the case for the sextet $  \Xi^{'}_b $ state,  preventing us to calculate the branching ratios for 
 $ \Xi^{'}_b\to\Xi^{'}_c   \ell {\overline{\nu}}_\ell$ and $  \Xi^{'}_b\to\Xi_c   \ell {\overline{\nu}}_\ell$ channels. For these channels, we only report the width of the transitions at different lepton channels.   In  Tables \ref{DECAY2},  \ref{DECAY3} and \ref{DECAY4}, we not only present the widths and branching ratios,  but also depict the ratios of branching fractions. In the case of $  \Xi^{'}_b $ particle in the initial states, these ratios are computed using the ratios of widths at different lepton channels.  Our predictions for different observables corresponding to different semileptonic weak channels can be compared with future theoretical estimations  and experimental results.

\begin{table}[h!]
\caption{Decay widths,  branching ratios and $R_{\Xi^{'}_c}$  for $ \Xi_b\to\Xi^{'}_c $ at different channels.}\label{DECAY2}
\begin{ruledtabular}
\begin{tabular}{|c|c|c|c|}
                 &$\Xi_b\to\Xi^{'}_c e {\overline{\nu}}_{e}$ &    $\Xi_b\to\Xi^{'}_c \mu {\overline{\nu}}_{\mu}$ & $\Xi_b\to\Xi^{'}_c\tau {\overline{\nu}}_{\tau}$\\
\hline
$\Gamma \times10^{14}\mathrm{GeV} $ &$2.48^{+1.08}_{-0.87}$ & $2.44^{+1.07}_{-0.87}$ & $0.44^{+0.21}_{-0.16}$\\
\hline
Br(\%) &$5.91^{+2.58}_{-2.10} $& $5.83^{+2.57}_{-2.07}$ & $1.05^{+0.49}_{-0.38}$\\
\hline
$R_{\Xi^{'}_c}=\frac{Br[\Xi_b\rightarrow \Xi^{'}_c\tau\overline{\nu}_\tau]}{Br[\Xi_b\rightarrow \Xi^{'}_c e \overline{\nu}_{e}]}$ $= 0.18 \pm 0.08$\\
\end{tabular}
\end{ruledtabular}
\end{table}

\begin{table}[h!]
\caption{Decay widths and $R_{\Xi^{'}_c}$  for $ \Xi^{'}_b\to\Xi^{'}_c $ at different channels.}\label{DECAY3}
\begin{ruledtabular}
\begin{tabular}{|c|c|c|c|}
                 &$\Xi^{'}_b\to\Xi^{'}_c e {\overline{\nu}}_{e}$ &    $\Xi^{'}_b\to\Xi^{'}_c \mu {\overline{\nu}}_{\mu}$ & $\Xi^{'}_b\to\Xi^{'}_c\tau {\overline{\nu}}_{\tau}$\\
\hline
$\Gamma \times10^{15}\mathrm{GeV} $ & $5.01^{+1.87}_{-1.50}$ & $4.92^{+1.85}_{-1.49}$ & $0.61^{+0.35}_{-0.24}$\\
\hline
$R_{\Xi^{'}_c}=\frac{\Gamma[\Xi^{'}_b\rightarrow \Xi^{'}_c\tau\overline{\nu}_\tau]}{\Gamma[\Xi^{'}_b\rightarrow \Xi^{'}_c e \overline{\nu}_{e}]}$ $=0.12\pm 0.05$\\
\end{tabular}
\end{ruledtabular}
\end{table}
 
\begin{table}[h!]
\caption{Decay widths and $R_{\Xi_c}$ for $ \Xi^{'}_b\to\Xi_c $ at different channels.}\label{DECAY4}
\begin{ruledtabular}
\begin{tabular}{|c|c|c|c|}
                 &$\Xi^{'}_b\to\Xi_c e {\overline{\nu}}_{e}$ &    $\Xi^{'}_b\to\Xi_c \mu {\overline{\nu}}_{\mu}$ & $\Xi^{'}_b\to\Xi_c\tau {\overline{\nu}}_{\tau}$ \\
\hline
$\Gamma \times10^{15}\mathrm{GeV} $ & $3.33^{+1.73}_{-1.38}$ & $3.29^{+1.73}_{-1.38}$ & $0.75^{+0.39}_{-0.31}$\\
\hline
$R_{\Xi_c}=\frac{\Gamma[\Xi^{'}_b\rightarrow \Xi_c\tau\overline{\nu}_\tau]}{\Gamma[\Xi^{'}_b\rightarrow \Xi_c e \overline{\nu}_{e}]}$ $=0.23\pm0.10$\\
\end{tabular}
\end{ruledtabular}
\end{table}

\section{Conclusion}~\label{Sec5}
Despite being the most successful framework for describing fundamental interactions, the SM has some limitations.  Recent discrepancies between the model's predictions and experimental observations especially at $ B $ meson decay channels  have motivated researchers to explore more decay modes, especially at baryonic channels.  The study of heavy b-baryon decays,  such as  $\Xi^{(')}_{b}\rightarrow \Xi^{(')}_{c}{\ell}\bar\nu_{\ell}$,  can provide valuable insights to test the SM predictions and  search for physics BSM. These specific decays have not yet been observed experimentally.  Our results may guide corresponding experimental activities for measuring different observables defining such decay channels.

We calculated the weak semileptonic form factors representing the low energy matrix elements of the decay modes $\Xi^{(')}_{b}\rightarrow \Xi^{(')}_{c}{\ell}\bar\nu_{\ell}$. Using these form factors, we computed the widths, branching ratios and ratio of branching fractions at different lepton channels.  Although we compared our results for some quantities with the existing literature, we provided many information to be compared with future theoretical estimations and experimental results.  Hence,  most of the information was provided for the first time in the literature.  Any comparison of our predictions with future possible experimental data is crucial for our understanding of not only the internal structure and properties of the participating particles, but for test of the SM predictions in b-baryon decays.

\section*{ACKNOWLEDGMENTS} 

K. Azizi  is grateful  to Iran national science foundation (INSF) for the partial financial support provided under the elites Grant No. 4037888. He  also thanks   the CERN-Theory department for their support and warm hospitality.


\appendix 

\section{The correlation function of other decay channels}
In this appendix, we provide the correlation function for other decay channels just after contraction of quark fields and in terms of the heavy and light quark propagators. The QCD correlation function for $\Xi^{-}_{b}\rightarrow
\Xi^{0'}_{c}{\ell}\bar\nu_{\ell}$ decay is:
\begin{eqnarray} \label{24 term}
&&\Pi^{QCD}_{\mu}(p,p',q)=i^2 \int d^4x e^{-ipx}\int d^4y e^{ip'y} \frac{-1}{\sqrt{12}} \epsilon_{a'b'c'} \epsilon_{abc}\Bigg\{-2\gamma_5 S^{cb'}_s(y-x) S'^{aa'}_d(y-x) S^{bi}_c(y) \gamma_\mu(1-\gamma_5) S^{ic'}(-x) \gamma_5\notag\\
&&+2\beta\gamma_5 S^{cb'}_s(y-x)  \gamma_5 S'^{aa'}_d(y-x) S^{bi}_c(y) \gamma_\mu(1-\gamma_5) S^{ic'}(-x)+Tr[S^{ib'}_b(-x) S'^{aa'}_d(y-x) S^{bi}_c(y) \gamma_\mu(1-\gamma_5)] \gamma_5 S^{cc'}_s(y-x) \gamma_5\notag\\
&&-\beta Tr[S^{ib'}_b(-x) \gamma_5 S'^{aa'}_d(y-x) S^{bi}_c(y) \gamma_\mu(1-\gamma_5)] \gamma_5 S^{cc'}_s(y-x)-\gamma_5 S^{cb'}_s(y-x) S'^{ia'}_b(-x) (1-\gamma_5)\gamma_\mu S'^{bi}_b(y) S^{ac'}_d(y-x)\gamma_5\notag\\
&&+\beta \gamma_5 S^{cb'}_s(y-x)\gamma_5 S'^{ia'}_b(-x) (1-\gamma_5)\gamma_\mu S'^{bi}_b(y) S^{ac'}_d(y-x)-2\beta S^{cb'}_s(y-x) S'^{aa'}_d(y-x)\gamma_5 S^{bi}_c(y)\gamma_\mu(1-\gamma_5) S^{ic'}_b(-x)\gamma_5\notag\\
&&+2\beta^2 S^{cb'}_s(y-x) \gamma_5 S'^{aa'}_d(y-x)\gamma_5 S^{bi}_c(y)\gamma_\mu(1-\gamma_5) S^{ic'}_b(-x)+\beta Tr[S'^{aa'}_d(y-x) \gamma_5 S^{bi}_c(y) \gamma_\mu(1-\gamma_5) S^{ib'}_b(-x)] S^{cc'}_s(y-x) \gamma_5\notag\\
&&-\beta^2 Tr[\gamma_5 S'^{aa'}_d(y-x) \gamma_5 S^{bi}_c(y) \gamma_\mu(1-\gamma_5) S^{ib'}_b(-x)] S^{cc'}_s(y-x)-\beta S^{cb'}_s(y-x) S'^{ia'}_b(-x) (1-\gamma_5)\gamma_\mu S'^{bi}_c(y)\gamma_5 S^{ac'}_d(y-x) \gamma_5\notag\\
&&+\beta^2 S^{cb'}_s(y-x) \gamma_5 S'^{ia'}_b(-x) (1-\gamma_5)\gamma_\mu S'^{bi}_c(y)\gamma_5 S^{ac'}_d(y-x)+2\gamma_5 S^{ca'}_d(y-x) S'^{bb'}_s(y-x) S^{ai}_c(y)\gamma_\mu(1-\gamma_5) S^{ic'}_b(-x) \gamma_5\notag\\
&&-2\beta \gamma_5 S^{ca'}_d(y-x) \gamma_5 S'^{bb'}_s(y-x) S^{ai}_c(y)\gamma_\mu(1-\gamma_5) S^{ic'}_b(-x)+\gamma_5 S ^{ca'}_d(y-x) S'^{ib'}_b(-x) (1-\gamma_5)\gamma_\mu S'^{ai}_c(y) S^{bc'}_s(y-x) \gamma_5\notag\\
&&-\beta \gamma_5 S ^{ca'}_d(y-x) \gamma_5 S'^{ib'}_b(-x) (1-\gamma_5)\gamma_\mu S'^{ai}_c(y) S^{bc'}_s(y-x)- Tr[S'^{ia'}_b(-x) (1-\gamma_5)\gamma_\mu S'^{ai}_c(y) S^{bb'}_s(y-x)] \gamma_5 S^{cc'}_d(y-x) \gamma_5\notag\\
&&+ Tr[\gamma_5 S'^{ia'}_b(-x) (1-\gamma_5)\gamma_\mu S'^{ai}_c(y) S^{bb'}_s(y-x)] \gamma_5 S^{cc'}_d(y-x)+2\beta S^{ca'}_d(y-x) S'^{bb'}_s(y-x) \gamma_5 S^{ai}_c(y) \gamma_\mu(1-\gamma_5) S^{ic'}_b(-x) \gamma_5\notag\\
&&-2\beta^2 S^{ca'}_d(y-x) \gamma_5 S'^{bb'}_s(y-x) \gamma_5 S^{ai}_c(y) \gamma_\mu(1-\gamma_5) S^{ic'}_b(-x)+\beta S^{ca'}_d(y-x) S'^{ib'}_b(-x) (1-\gamma_5)\gamma_\mu S'^{ai}_c(y)\gamma_5 S^{bc'}_s(y-x) \gamma_5\notag\\
&&-\beta^2 S^{ca'}_d(y-x) \gamma_5 S'^{ib'}_b(-x) (1-\gamma_5)\gamma_\mu S'^{ai}_c(y)\gamma_5 S^{bc'}_s(y-x)-\beta Tr[S'^{ia'}_b(-x)(1-\gamma_5)\gamma_\mu S'^{ai}_c(y)\gamma_5 S^{bb'}_s(y-x)] S^{cc'}_d(y-x) \gamma_5\notag\\
&&+\beta^2 Tr[\gamma_5 S'^{ia'}_b(-x)(1-\gamma_5)\gamma_\mu S'^{ai}_c(y)\gamma_5 S^{bb'}_s(y-x)] S^{cc'}_d(y-x)
\Bigg\}.
\end{eqnarray}

For the QCD correlation function of $\Xi^{-'}_{b}\rightarrow\Xi^{0'}_{c}{\ell}\bar\nu_{\ell}$ decay, we obtain:
\begin{eqnarray} \label{24 term}
&&\Pi^{QCD}_{\mu}(p,p',q)=i^2 \int d^4x e^{-ipx}\int d^4y e^{ip'y} \frac{1}{2} \epsilon_{a'b'c'} \epsilon_{abc}\Bigg\{Tr[S'^{aa'}_d(y-x) S^{bi}_c(y) \gamma_\mu(1-\gamma_5) S^{ib'}_b(-x)] \gamma_5 S^{cc'}_s(y-x) \gamma_5\notag\\
&&-\beta Tr[\gamma_5 S'^{aa'}_d(y-x) S^{bi}_c(y) \gamma_\mu(1-\gamma_5) S^{ib'}_b(-x)] \gamma_5 S^{cc'}_s(y-x) +\gamma_5 S^{cb'}_s(y-x) S'{ia'}_b(-x) (1-\gamma_5)\gamma_\mu S'^{bi}_c(y) S^{ac'}_d(y-x)\gamma_5\notag\\
&&-\beta \gamma_5 S^{cb'}_s(y-x) \gamma_5 S'{ia'}_b(-x) (1-\gamma_5)\gamma_\mu S'^{bi}_c(y) S^{ac'}_d(y-x)+\beta Tr[S'^{aa'}_d(y-x) \gamma_5 S^{bi}_c(y)\gamma_\mu(1-\gamma_5) S^{ib'}_b(-x)] S^{cc'}_s(y-x)\gamma_5\notag\\
&&-\beta^2 Tr[\gamma_5 S'^{aa'}_d(y-x) \gamma_5 S^{bi}_c(y)\gamma_\mu(1-\gamma_5) S^{ib'}_b(-x)] S^{cc'}_s(y-x)+\beta S^{cb'}_s(y-x) S'^{ia'}_b(-x) (1-\gamma_5)\gamma_\mu S'^{bi}_c(y)\gamma_5 S^{ac'}_d(y-x)\gamma_5\notag\\
&&-\beta^2 S^{cb'}_s(y-x) \gamma_5 S'^{ia'}_b(-x) (1-\gamma_5)\gamma_\mu S'^{bi}_c(y)\gamma_5 S^{ac'}_d(y-x)+\gamma_5 S^{ca'}_d(y-x) S'^{ib'}_b(-x) (1-\gamma_5)\gamma_\mu S'^{ai}_c(y) S^{bc'}_s(y-x)\gamma_5\notag\\
&&-\beta\gamma_5 S^{ca'}_d (y-x)\gamma_5 S'^{ib'}_b(-x) (1-\gamma_5)\gamma_\mu S'^{ai}_c(y) S^{bc'}_s(y-x)+Tr[S'^{ia'}_b(-x)(1-\gamma_5)\gamma_\mu S'^{ai}_c(y) S^{bb'}_s(y-x)] \gamma_5 S^{cc'}_d(y-x)\gamma_5\notag\\
&&-\beta Tr[\gamma_5 S'^{ia'}_b(-x)(1-\gamma_5)\gamma_\mu S'^{ai}_c(y) S^{bb'}_s(y-x)] \gamma_5 S^{cc'}_d(y-x)+\beta S^{ca'}_d(y-x) S'^{ib'}_b(-x)(1-\gamma_5)\gamma_\mu S'^{ai}_c(y)\gamma_5 S^{bc'}_s(y-x)\gamma_5\notag\\
&&-\beta^2 S^{ca'}_d(y-x) \gamma_5 S'^{ib'}_b(-x)(1-\gamma_5)\gamma_\mu S'^{ai}_c(y)\gamma_5 S^{bc'}_s(y-x)+\beta Tr[S'^{ia'}_b(-x)(1-\gamma_5)\gamma_\mu S'^{ai}_c(y) \gamma_5 S^{bb'}_s(y-x)] S^{cc'}_d(y-x)\gamma_5\notag\\
&&-\beta^2 Tr[\gamma_5 S'^{ia'}_b(-x)(1-\gamma_5)\gamma_\mu S'^{ai}_c(y) \gamma_5 S^{bb'}_s(y-x)] S^{cc'}_d(y-x)
\Bigg\}. 
\end{eqnarray}
And,  the QCD correlation function for $\Xi^{-'}_{b}\rightarrow\Xi^{0}_{c}{\ell}\bar\nu_{\ell}$ decay is found to be:
\begin{eqnarray} \label{24 term}
&&\Pi^{QCD}_{\mu}(p,p',q)=i^2 \int d^4x e^{-ipx}\int d^4y e^{ip'y} \frac{-1}{\sqrt{12}} \epsilon_{a'b'c'} \epsilon_{abc}\Bigg\{-2\gamma_5 S^{ci}_c(y) \gamma_\mu(1-\gamma_5) S^{ib'}_b(-x) S'^{aa'}_d(y-x) S^{bc'}_s(y-x)\gamma_5\notag\\
&&+2\beta \gamma_5 S^{ci}_c(y) \gamma_\mu(1-\gamma_5) S^{ib'}_b(-x) \gamma_5 S'^{aa'}_d(y-x) S^{bc'}_s(y-x)+2\gamma_5 S^{ci}_c(y)\gamma_\mu(1-\gamma_5) S^{ia'}_b(-x) S'^{bb'}_s(y-x) S^{ac'}_d(y-x)\gamma_5\notag\\
&&-2\beta \gamma_5 S^{ci}_c(y)\gamma_\mu(1-\gamma_5) S^{ia'}_b(-x) \gamma_5 S'^{bb'}_s(y-x) S^{ac'}_d(y-x)-2\beta S^{ci}_c(y)\gamma_\mu(1-\gamma_5) S^{ib'}_b(-x) S'^{aa'}_d(y-x)\gamma_5 S^{bc'}_s(y-x)\gamma_5\notag\\
&&+2\beta^2 S^{ci}_c(y)\gamma_\mu(1-\gamma_5) S^{ib'}_b(-x)\gamma_5 S'^{aa'}_d(y-x)\gamma_5 S^{bc'}_s(y-x)+2\beta S^{ci}_c(y)\gamma_\mu(1-\gamma_5) S^{ia'}_b(-x)S'^{bb'}_s(y-x)\gamma_5 S^{ac'}_d(y-x)\gamma_5\notag\\
&&-2\beta^2  S^{ci}_c(y)\gamma_\mu(1-\gamma_5) S^{ia'}_b(-x)\gamma_5S'^{bb'}_s(y-x)\gamma_5 S^{ac'}_d(y-x)+Tr[S'^{aa'}_d(y-x) S^{bi}_c(y)\gamma_\mu(1-\gamma_5) S^{ib'}_b(-x)] \gamma_5 S^{cc'}_s(y-x)\gamma_5\notag\\
&&-\beta Tr[S'^{aa'}_d(y-x) S^{bi}_c(y)\gamma_\mu(1-\gamma_5) S^{ib'}_b(-x)\gamma_5] \gamma_5 S^{cc'}_s(y-x)+\gamma_5 S^{cb'}_s(y-x) S'^{ia'}_b(-x)(1-\gamma_5)\gamma_\mu S'^{bi}_c(y) S^{ac'}_d(y-x)\gamma_5\notag\\
&&-\beta \gamma_5 S^{cb'}_s(y-x) \gamma_5 S'^{ia'}_b(-x)(1-\gamma_5)\gamma_\mu S'^{bi}_c(y) S^{ac'}_d(y-x)+\beta Tr[S'^{aa'}_d(y-x)\gamma_5 S^{bi}_c(y) \gamma_\mu(1-\gamma_5) S^{ib'}_b(-x)]S^{cc'}_s(y-x)\gamma_5\notag\\
&&-\beta^2Tr[S'^{aa'}_d(y-x)\gamma_5 S^{bi}_c(y) \gamma_\mu(1-\gamma_5) S^{ib'}_b(-x)\gamma_5]S^{cc'}_s(y-x)+ \beta S^{cb'}_s(y-x) S'^{ia'}_b(-x)(1-\gamma_5)\gamma_\mu S'^{bi}_c(y)\gamma_5 S^{ac'}_d(y-x)\gamma_5\notag\\
&&-\beta^2 S^{cb'}_s(y-x)\gamma_5 S'^{ia'}_b(-x)(1-\gamma_5)\gamma_\mu S'^{bi}_c(y)\gamma_5 S^{ac'}_d(y-x)-\gamma_5 S^{ca'}_d(y-x) S'^{ib'}_b(-x)(1-\gamma_5)\gamma_\mu S'^{ai}_c(y) S^{bc'}_s(y-x)\gamma_5\notag\\
&&+\beta \gamma_5 S^{ca'}_d(y-x) \gamma_5S'^{ib'}_b(-x)(1-\gamma_5)\gamma_\mu S'^{ai}_c(y) S^{bc'}_s(y-x)-Tr[S'^{bb'}_s(y-x) S^{ai}_c(y)\gamma_\mu(1-\gamma_5) S^{ia'}_b(-x)]\gamma_5S^{cc'}_d(y-x)\gamma_5\notag\\
&&+\beta Tr[S'^{bb'}_s(y-x) S^{ai}_c(y)\gamma_\mu(1-\gamma_5) S^{ia'}_b(-x)\gamma_5]\gamma_5S^{cc'}_d(y-x)-\beta S^{ca'}_d(y-x) S'^{ib'}_b(-x)(1-\gamma_5)\gamma_\mu S'^{ai}_c(y)\gamma_5 S^{bc'}_s(y-x)\gamma_5\notag\\
&&+\beta^2 S^{ca'}_d(y-x) \gamma_5 S'^{ib'}_b(-x)(1-\gamma_5)\gamma_\mu S'^{ai}_c(y)\gamma_5 S^{bc'}_s(y-x)-2\beta Tr[S^{ia'}_b(-x) S'^{bb'}_s(y-x)\gamma_5 S^{ai}_c(y)\gamma_\mu(1-\gamma_5)]S^{cc'}_d(y-x)\gamma_5\notag\\
&&+\beta^2  Tr[S^{ia'}_b(-x)\gamma_5 S'^{bb'}_s(y-x)\gamma_5 S^{ai}_c(y)\gamma_\mu(1-\gamma_5)]S^{cc'}_d(y-x)
\Bigg\}.
\end{eqnarray}

\section{Some  details of the calculations  in QCD side}
After substituting the quark propagators into the correlation function, the terms appear as follows: 
\begin{eqnarray}\label{exampleterm}
\int d^4k\int d^4k' \int d^4x e^{i(k-p).x}\int d^4y e^{i(-k'+p').y} \frac{x_\mu y_\nu k_{\mu'} k'_{\nu'}}{(k^2-m_b^2)(k'^2-m_c^2)[(y-x)^2]^n}.\notag\\
\end{eqnarray}
Using the identity below,  $x$ and $y$ appear in the  exponential form \cite{Azizi:2017ubq}:

\begin{eqnarray}\label{intyx}
\frac{1}{[(y-x)^2]^n}&=&\int\frac{d^Dt}{(2\pi)^D}e^{-it\cdot(y-x)}~i~(-1)^{n+1}~2^{D-2n}~\pi^{D/2}  \frac{\Gamma(D/2-n)}{\Gamma(n)}\Big(-\frac{1}{t^2}\Big)^{D/2-n},
\end{eqnarray}
by substituting  $x_{\mu}\rightarrow
i\frac{\partial}{\partial p_{\mu}}$ and $y_{\mu}\rightarrow
-i\frac{\partial}{\partial p'_{\mu}}$,  we get:

\begin{eqnarray}\label{expyx}
\int d^Dt \int d^4k\int d^4k' \int d^4x e^{i(k-p+t).x}\int d^4y e^{i(-k'+p'-t).y}  \frac{f(k,k',t)}{(k^2-m_b^2)(k'^2-m_c^2)}.
\end{eqnarray}
Now the Fourier integrals can be performed:
\begin{eqnarray}\label{fourier}
 \int d^4x e^{i(k-p+t).x}\int d^4ye^{i(-k'+p'-t).y}= (2\pi)^4\delta^4(k-p+t) (2\pi)^4\delta^4(-k'+p'-t).
\end{eqnarray}
The Dirac deltas remove the integrals over $k$ and $k'$ and only a D-dimensional integral over $t$ remains, which is calculated using Feynman parametrization and \cite{Azizi:2017ubq}:
%
\begin{eqnarray}\label{Int}
\int d^Dt\frac{(t^2)^{m}}{(t^2+L)^{n}}=\frac{i \pi^2
(-1)^{m-n}\Gamma(m+2)\Gamma(n-m-2)}{\Gamma(2)
\Gamma(n)[-L]^{n-m-2}}.\quad
\end{eqnarray}
The appearing gamma functions are transformed into logaritmic and imaginary functions using the identity below \cite{Azizi:2017ubq}:
\begin{eqnarray}\label{gamma}
\Gamma[\frac{D}{2}-n](-\frac{1}{L})^{D/2-n}=\frac{(-1)^{n-1}}{(n-2)!}(-L)^{n-2}ln[-L].
\end{eqnarray}

\section{Perturbative and non-perturbative contributions}
 The explicit forms for the perturbative and non-perturbative components  for the structure $p_{\mu}\slashed{p}'\slashed{p}$ in the channel $ \Xi_b\to\Xi_c \ell {\overline{\nu}}_\ell $, for example,  are given as:
\begin{eqnarray} \label{RhoPert}
&\rho^{Pert.}_{p_{\mu}\slashed{p}'\slashed{p}}(s,s',q^2)=\int_{0}^{1}du \int_{0}^{1-u}dv~\frac{1}{768 \pi^4 Z^2}
\Bigg\{ (3+\beta) m_c u \Big (u^2+ (-1+v) v\Big) + v Z \notag\\
& \Big [3\beta m_s  (1+6u-v)+ \beta m_b v+ 3 \beta^2 m_s (-1+5u+v) +3(m_s u+m_b v)+\notag\\
&3 m_d\Big (-1+2 u+\beta(4+3 u-4v)+v+3\beta^2 Z\Big)\Big]\Bigg\} ~\Theta[\Delta(s,s^{\prime},q^2)],
\end{eqnarray}

\begin{eqnarray} \label{Rho3}
\rho^3_{p_{\mu}\slashed{p}'\slashed{p}}(s,s',q^2)=0,
\end{eqnarray}
\begin{equation}\label{Rho4}
\rho^4_{p_{\mu}\slashed{p}'\slashed{p}}(s,s',q^2)=0,
\end{equation}

and
\begin{eqnarray}\label{gamma5,6}
\Gamma_{p_{\mu}\slashed{p}'\slashed{p}}(p^2,p'^2,q^2)=\Gamma^5_{p_{\mu}\slashed{p}'\slashed{p}}(p^2,p'^2,q^2)+\Gamma^6_{p_{\mu}\slashed{p}'\slashed{p}}(p^2,p'^2,q^2),
\end{eqnarray}
with
\begin{eqnarray}\label{Rho5}
&&\Gamma^5_{p_{\mu}\slashed{p}'\slashed{p}}(p^2,p'^2,q^2)=
\frac{(-1+\beta) m_o^2 m_s}{1152 \pi^2 r^2 r'}\Big[
\beta m_s \langle \bar{d}d\rangle-m_d \langle \bar{s}s\rangle+3\beta m_d\langle \bar{s}s\rangle\Big],
\end{eqnarray}
and
\begin{eqnarray}\label{Rho6}
&&\Gamma^6_{p_{\mu}\slashed{p}'\slashed{p}}=\frac{1}{7776 \pi^2 r^3 r'^2} \Bigg\{ \langle \bar{s}s\rangle \Big [ 27 \langle \bar{d}d\rangle m_s \pi^2 (3 m_c m_d r+3 m_b m_d r'-2r r')+g_s^2 m_d r r' \langle \bar{s}s\rangle \Big] \notag\\
&&+\beta^2 r r'\Big( \langle \bar{d}d\rangle ^2 g_s^2 m_s-54 \langle \bar{d}d\rangle (m_d+3m_s) \pi^2 \langle \bar{s}s\rangle+3 g_s^2 m_d \langle \bar{s}s\rangle^2\Big)+\beta\Big[- \langle \bar{d}d\rangle ^2 g_s^2 m_s r r'\notag\\
&&+ 27 \langle \bar{d}d\rangle  \pi^2 \Big (m_c m_d m_s r+m_b m_d m_s r'+2(m_d+4m_s) r r'\Big ) \langle \bar{s}s\rangle -4g_s^2 m_d r r'  \langle \bar{s}s\rangle^2\Big]\Bigg\}.
\end{eqnarray}
In the above equations we have used the following short-hand notations:
\begin{eqnarray}\label{L}
&&\Delta(s,s^{\prime},q^2)=- m_c^2 u + s^{\prime} u - s^{\prime}
u^2 - m_b^2 v + s v+ q^2 u v - s u v - s^{\prime} u v - s v^2,\notag
\\
&&Z=u+v-1,\notag
\\
&&r=m_b^2-p^2,\notag
\\
&&r'=m_c^2-p'^2,\notag
\\
\end{eqnarray}
and $\Theta[...]$ stands for the unit-step function.

\label{sec:Num}

\end{document}